\documentclass[3p,times]{elsarticle}

\usepackage{ecrc}

\usepackage{subfig}
\usepackage{floatrow}
\usepackage{amsmath}

\volume{00}

\firstpage{1}

\journalname{Nuclear Instrumentation and Methods A}

\runauth{J. Adams et al.}

\jid{NIMA}

\jnltitlelogo{Nucl. Instr. Meth. A}

\usepackage{amssymb}
\usepackage{url}

\begin{document}

\begin{frontmatter}




\title{The STAR Event Plane Detector}


\author[1]{J.~Adams}
\author[2]{A.~Ewigleben}
\author[3]{S.~Garrett\footnote{Currently Facility for Rare Isotope Beams, Michigan State University, East Lansing Michigan 48824.}}
\author[4]{W.~He}
\author[5]{T.~Huang}
\author[3]{P.M.~Jacobs}
\author[6]{X.~Ju}
\author[1]{M.A.~Lisa\corref{cor1}}
\author[3,7]{M.~Lomnitz}
\author[8]{R.~Pak}
\author[2]{R.~Reed}
\author[3]{A.~Schmah\footnote{Currently Universit{\"a}t Heidelberg, Heidelberg 69120}}
\author[2]{P.~Shanmuganathan\footnote{Currently Brookhaven National Laboratory, Upton, New York 11973.}}
\author[6]{M.~Shao}
\author[3]{X.~Sun\footnote{Currently University of Illinois, Chicago Illinois 60607}}
\author[1]{I.~Upsal\footnote{Currently Brookhaven National Laboratory, Upton, New York 11973, and Shandong University, Qingdao, Shandong 266237}}
\author[9]{G.~Visser}
\author[3]{J.~Zhang}

\cortext[cor1]{Corresponding author.  lisa.1@osu.edu}
\address[1]{The Ohio State University, Columbus, Ohio 43210}
\address[2]{Lehigh University, Bethlehem, Pennsylvania 18015}
\address[3]{Lawrence Berkeley National Laboratory, Berkeley, California 94720}
\address[4]{Fudan University, Shanghai 200433}
\address[5]{National Cheng Kung University, Tainan 70101}
\address[6]{University of Science and Technology of China, Hefei, Anhui 230026}
\address[7]{Kent State University, Kent, Ohio 44242}
\address[8]{Brookhaven National Laboratory, Upton, New York 11973}
\address[9]{Indiana University, Bloomington, Indiana 47408}

%%\author{}
%%\address{}

\begin{abstract}
The Event Plane Detector (EPD) is an upgrade detector to the STAR experiment at RHIC,
  designed to measure the pattern of forward-going charged particles emitted in a
  high-energy collision between heavy nuclei.
It consists of two highly-segmented disks of 1.2-cm-thick scintillator embedded with
  wavelength-shifting fiber, coupled to silicon photomultipliers and custom electronics.
We describe the general design of the device, its construction, and performance on the
  bench and in the experiment.
\end{abstract}


\end{frontmatter}

\setcounter{footnote}{0}

%%\linenumbers

%%%%%%%%%%%%%%%%%%%%%%%%% Section 1 - Introduction

\section{Introduction}
\label{sec:Sintroduction}

Collisions between heavy nuclei at high energies produce the quark-gluon plasma (QGP), a new
  state of matter consisting of deconfined quarks and gluons last seen only microseconds after
  the Big Bang~\cite{Bjorken:1982qr,Adams:2005dq}.
Determining the initial geometry of each collision is a crucial step in decoding the complex
  physics of this novel system~\cite{Heinz:2013th,Voloshin:2008dg}.
Experimentally, this geometry is quantified in terms of ``event planes'' (EPs), and there
  exist well-established methods~\cite{Poskanzer:1998yz,Richardson:2010hm} to extract these planes from
  patterns in the thousands of particles emitted from each event.
  
The STAR experiment~\cite{Ackermann:2002ad} at Brookhaven National Laboratory's Relativistic Heavy Ion
  Collider (RHIC)
  facility measures these collisions with a coordinated set of tracking, calorimetric, time-of-flight
  and other detector systems.\footnote{\url{https://www.star.bnl.gov/}}
These components form nearly hermetic coverage of the collision generated when two opposing beams
  of nuclei collide at the center of the experiment.
The Event Plane Detector measures charged particles emitted in the forward and backward
  directions, at angles $0.7^\circ<\theta<13.5^\circ$ relative to the initial directions of the 
  beams.
(Expressed in pseudorapidity, $\eta\equiv-\ln\left[\tan\left(\theta/2\right)\right]$, the
  detector covers $2.14<|\eta|<5.09$.)

\begin{figure}[t]
\includegraphics[width=0.8\textwidth]{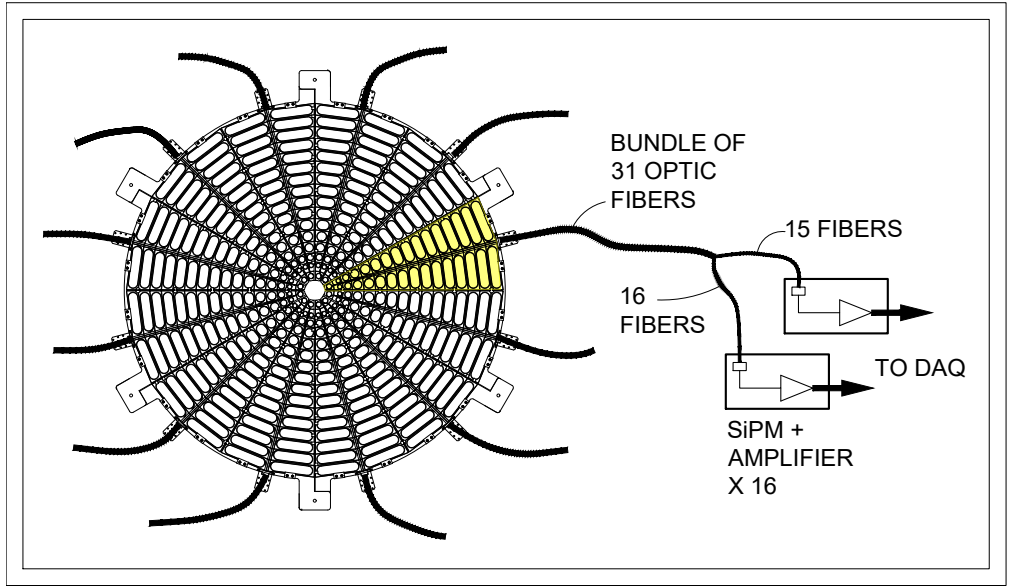}
\caption{A sketch of the EPD system.  One of two EPD wheels is shown.  The 31 tiles from each of 12 supersectors are connected via optical fiber bundles to silicon photomultipliers and amplification electronics.  See text for details.
}
\label{fig:Soverview}
\end{figure}

The EPD system consists of two wheels, one of which
 is shown schematically in figure~\ref{fig:Soverview}.
Each wheel is composed of 12 ``supersectors'' which subtend 30 degrees in azimuth, and which are divided into 31 tiles.
Each supersector is connected to a bundle of 31 optical fibers  which transport the light from each tile to a silicon
 photomultiplier (SiPM).
Signals from the SiPM are amplified and sent to the STAR digitizing and acquisition system~\cite{Landgraf:2002zw}.
In what follows, we discuss the design, construction, and bench tests of these components in detail.
We then discuss the system's performance as deployed in the STAR experiment and conclude.

%%%%%%%%%%%%%%%%%%% Section 2 - Design and Construction

\section{Design and Construction}
\label{sec:Swheels}

\subsection{General Considerations}
\label{sec:SwheelsGeneralDesignConsiderations}

\noindent{\bf Materials}\\
Event planes are extracted from the azimuthal (about the beam, or $z$ direction)
  patterns of emitted particles, and the species
  of the particles (protons, pions, etc.) is unimportant; therefore, no particle identification
  capability is required.
It is likewise not very important to know whether a detected particle originates directly
  from the collision vertex, or whether it is a daughter of a particle that decays in flight. 
Although the trajectories of charged particles are bent by STAR's 0.5~T magnetic field,
  our simulations showed that this bending does not significantly affect EP determination.
For these reasons, no tracking capability is required, and recording a single hit point for
  each particle is sufficient.

In addition to EP determination in offline analysis, the detector is used to trigger STAR's
  data acquisition system online.
This dictated a detector response time on order of nanoseconds.

These modest requirements prompted a design based on plastic scintillator (Eljen EJ-200) planes 1.2~cm thick.
Relativistic charged particles passing through such planes produce a signal distribution with identifiable peak features corresponding
  to 1, 2, 3, \ldots N minimally ionizing particles (MIPs) traversing the plastic simultaneously.  %% (though irreducible Landau fluctuations are significant).

This choice implies a relatively small addition to STAR's material budget (detailed below) and relatively low cost.
In addition, scintillator is straightforward to machine with CNC mills at university shops,
  and amenable to considerable manual construction by teams of students, providing
  valuable hardware experience to young members of the field. 

\noindent{\bf Geometry}\\
Because event planes are determined by Fourier decomposition of the azimuthal dependence of particle yields~\cite{Poskanzer:1998yz},
  it is best and most natural that an EPD have azimuthally symmetric geometry.
The spatial positions of other detector systems in STAR dictated that the EPD sit at
  $z=\pm3.75$~m, and coverage of pseudorapidity range $2.14<|\eta|<5.09$ is
  scientifically desired.
These considerations prompted the chosen geometry of a disk of radius 0.9~m,
  with an inner hole of radius 4.6~cm.
The vacuum pipe (radius 4~cm) carrying the nuclear beams passes through this inner hole.
The two beams travel east and west; one EPD wheel is located on the west side of the experiment,
  and one on the east.
  
It was decided to construct each wheel from 12 separate wedges, or ``supersectors''.
This decision was driven by several considerations.
In case of a major problem during machining (which occurred with two of the 32 wedges produced),
  only a limited number of tiles are ruined.
Also, the EPD is mounted every year onto STAR using manlifts in a tight spacing with delicate cables
  and vacuum tubes nearby; larger pieces are more awkward and prone to cause problems.
Finally, wedges on order 50~cm~$\times$~100~cm fit on standard CNC  mills available in university machine shops and are convenient to handle during manual
  stages of construction.

\noindent{\bf Segmentation}\\
Each disk was segmented into several tiles, each individually read out.
Several considerations argued for a small number of large tiles.
Naturally, for cost reasons and simplicity, the channel count should be kept as low as possible without degrading detector capabilities below experimental requirements.
Furthermore, increasing the number of tiles means decreasing their size.
Small tiles are difficult to work on manually, and it is difficult to respect the minimum bending radius of the wavelength shifting fibers used to transport light.
Many small tiles also means loss of efficiency in the inevitable dead areas
  between active tiles.

Scientific performance requirements lead to competing considerations.
The resolution with which an EP can be measured is ultimately limited by the physics of the nuclear collision itself;
  this limit is determined by models and previous measurements.
The azimuthal segmentation of the EPD should not be so coarse as to degrade the EP resolution beyond this limit.
In these collisions, the azimuthal pattern of particle emission is known to depend strongly on
  $\eta$~\cite{Back:2004zg,Back:2005pc}, and it is
  important to measure this dependence.
Therefore, sufficient segmentation in the radial direction is required.
Finally, due to irreducible Landau fluctuations, explained in section~\ref{sec:TestsPerformance},
  it is difficult to distinguish an event in which a single MIP
  passes through a tile, from one in which multiple MIPs traversed it simultaneously.
The multiple-MIP probability is driven by tile size and the distribution of particles emitted from the collision,
  which depends on $\eta$.
Based on previous measurements~\cite{Alver:2010ck}, the tiles sizes used in the EPD result in $\sim10\%$ double-hit
  probability for tiles at all radii, for central Au $+$ Au collisions at $\sqrt{s_{NN}}=19.6$~GeV.
The impact of multiple-hits on a specific offline analysis and online triggering must be determined from simulation.

Balancing these competing considerations resulted~\cite{Schmah:2020yxi} in a segmentation scheme, detailed below, in which each wheel consists of 372 tiles.

\subsection{Detailed design}
\label{sec:SdetailedWheelDesign}

Detailed dimensions and locations of the 
  EPD tile with respect to the detector's position within STAR are tabulated in table~\ref{table:EPD_Segmentation}. 

\begin{figure}[t]
  \includegraphics[width=70mm]{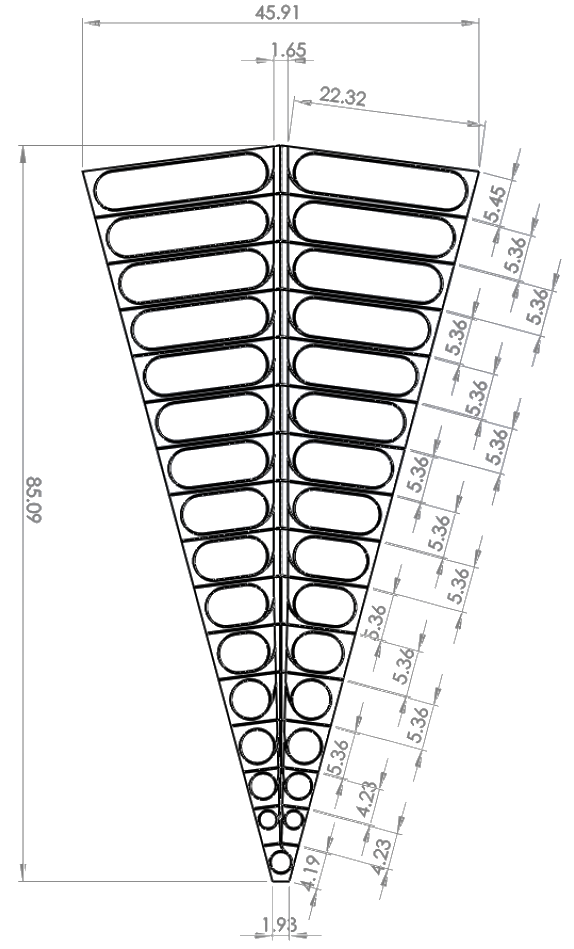}
  \caption{
  Detailed sketch of one supersector, with scales given in cm.  The central channel runs vertically in this figure, widening and deepening as it approaches the outer edge, to accommodate more WLS fibers.
  Straight isolation grooves, running approximately horizontal in this image, separate the tiles optically.
  Each tile contains a curved track groove near its periphery, into which a WLS fiber is glued.
  Except for at the tip (the part closest to the nuclear beam), each of the 16 ``rows'' contains two tiles per supersector.
  Dividing the tip (row 1) into two tiles would violate the minimum
  bending radius of the WLS fibers.
  \label{fig:SSdimensions}
}
\end{figure}

Supersectors were milled from 
  sheets of 1.2~cm thick scintillator (Eljen-EJ200).
The machine bit paths are shown in figure~\ref{fig:SSdimensions}.
Individual tiles are defined by 1.65-mm-wide ``isolation grooves'' that run the full depth of the
  scintillator.  
(These isolation grooves are machined in two stages, as discussed in section~\ref{sec:Construction}.)
Within each tile is a
  rounded groove track 1.6~mm wide and 3.6~mm deep.
These ``fiber grooves'' are filled with a wavelength 
  shifting (WLS) optical fiber (S-type Kuraray Y-11(200))
  wound three times around the groove, to optimize light
  collection then held in place by EJ-500 optical epoxy.
Our prototype tests indicated that triple winding of the fibers increases the light output of the detector by a factor of two over a single loop, in rough agreement with previous observations~\cite{Filippov:2000kma}.
  
Light collection of WLS fibers in similar channels has been reported~\cite{Filippov:2000kma} to be significantly better with epoxy, as compared with air coupling.
Especially in our case, using a groove with fixed width, epoxy was also crucial to keep the fibers from springing out of the groove.
The particular epoxy used (Eljen EJ-500) was extensively tested and found sufficiently radiation-hard for conditions at RHIC~\cite{Buechel:2017ofy}.

As they exit the grooved track of a tile, the WLS fibers are routed through a ``central channel'' 
  that runs down the center of the supersector.
The central channel increases in width and depth as it approaches the outer edge of the supersector. 
At this outer edge, the 31 individual fibers are bundled together in a ``32-fiber connector'' (one channel of which is empty). 
  The connector and fiber bundle that mates to it are described in section~\ref{sec:Sfibers}
  
\begin{table}[ht]
\centering
\caption{Columns 2 and 3 list minimum and maximum radial distance from the beamline for each row of tiles.   Columns 4 and 5 give minimum and maximum
pseudorapidity values corresponding to a particle traveling in a straight
line from the center of the STAR experiment, 375~cm in the beam direction.}
\label{table:EPD_Segmentation}
\begin{tabular}{|c|c|c|c|c|}
\hline
Row & ${r}_{i}$ (cm) & ${r}_{f}$ (cm) & ${\eta}_{i}$ & ${\eta}_{f}$ \\
\hline
1   & 4.6            & 9.0            & 5.09         & 4.42         \\
\hline
2   & 9.0            & 13.4           & 4.42         & 4.03         \\
\hline
3   & 13.4           & 17.8           & 4.03         & 3.74         \\
\hline
4   & 17.8           & 23.33          & 3.74         & 3.47         \\
\hline
5   & 23.33          & 28.86          & 3.47         & 3.26         \\
\hline
6   & 28.86          & 34.39          & 3.26         & 3.08         \\
\hline
7   & 34.39          & 39.92          & 3.08         & 2.94         \\
\hline
8   & 39.92          & 45.45          & 2.94         & 2.81         \\
\hline
9   & 45.45          & 50.98          & 2.81         & 2.69         \\
\hline
10  & 50.98          & 56.51          & 2.69         & 2.59         \\
\hline
11  & 56.51          & 62.05          & 2.59         & 2.50         \\
\hline
12  & 62.05          & 67.58          & 2.50         & 2.41         \\
\hline
13  & 67.58          & 73.11          & 2.41         & 2.34         \\
\hline
14  & 73.11          & 78.64          & 2.34         & 2.27         \\
\hline
15  & 78.64          & 84.17          & 2.27         & 2.20         \\
\hline
16  & 84.17          & 89.70          & 2.20         & 2.14         \\  
\hline
\end{tabular}
\end{table}

\subsection{Construction}
\label{sec:Construction}

Here, we detail the process for constructing a supersector.

The scintillator wedge was machined with a CNC mill in two stages, described below.
The toolpath included ramping fiber paths and isolation grooves rather deep (6~mm) compared to their width (1.6~mm).
A high-strength carbide cutter was run at a high spin rate  (8065 RPM) and low feed rate (3 in./minute).
A high flow rate of water and 5\% oil was maintained by adding coolant lines to the CNC machine, and 
  little material (30 mil) was removed with each tool pass.
These precautions prevented scintillator melting as well as cracking that can result when
  small fragments get caught between the bit and the grooves.
The large scintillator sheets were clamped for many hours during machining, and
  considerable care and experimentation was required to find a clamping scheme
  that avoided the micro-cracking which can easily occur with scintillator sheets under
  compression and vibration.

\begin{figure}[p]
\subfloat[Removing tape from the supersector edge.]
{\includegraphics[width=0.45\textwidth]{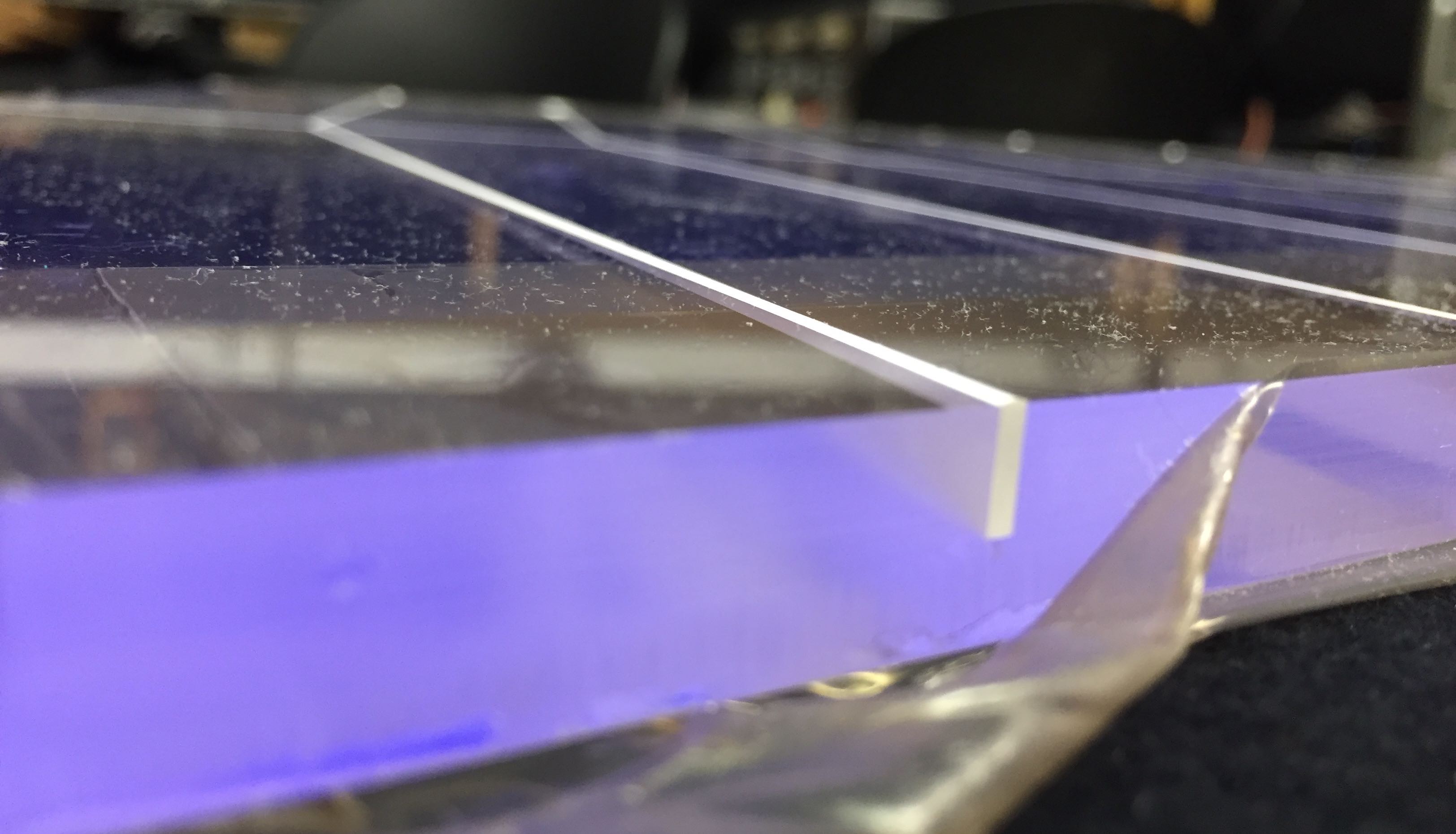}\label{fig:SedgeTape}}\\
\subfloat[After the second milling stage.]
{\includegraphics[width=0.3\textwidth]{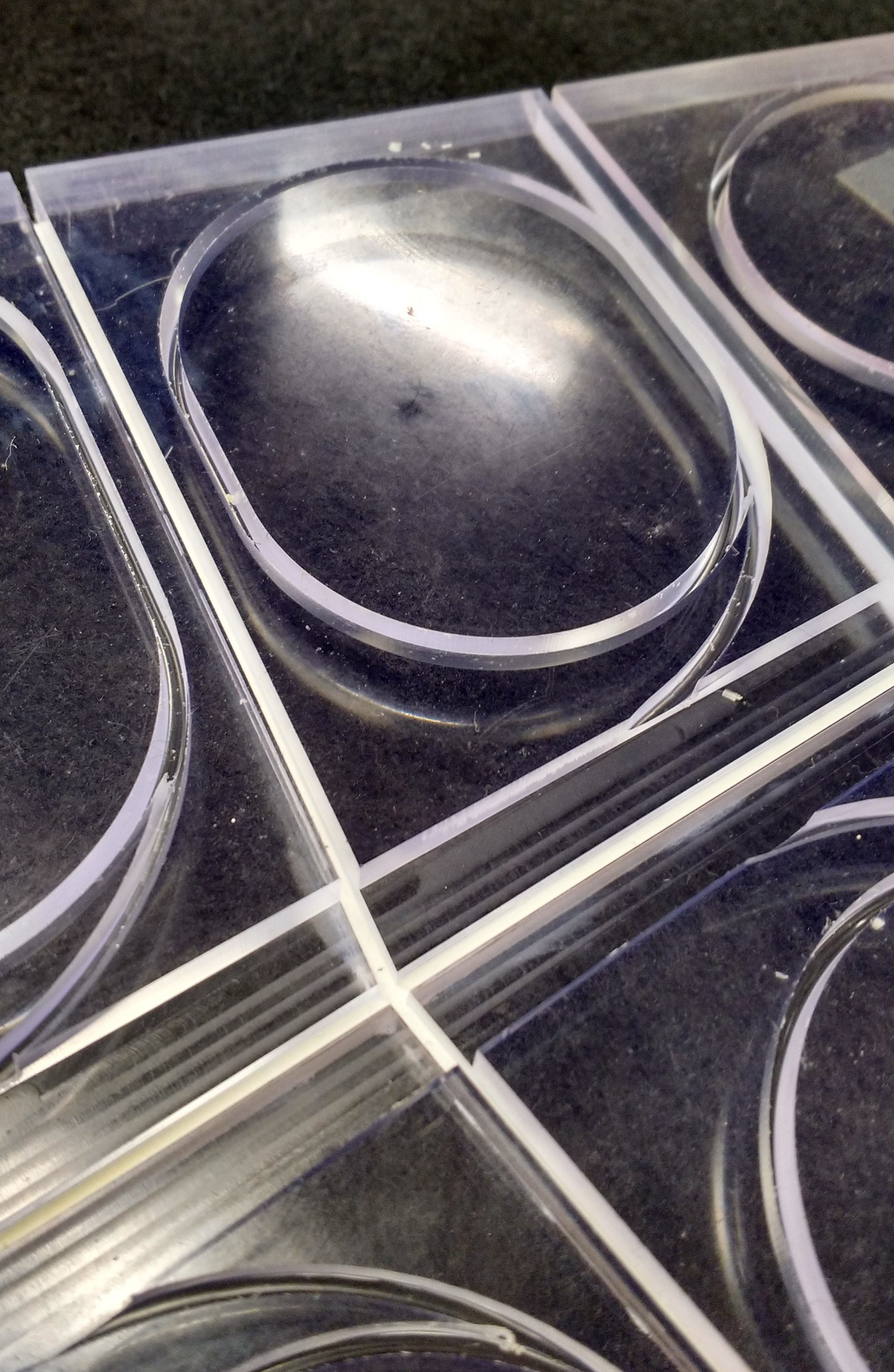}\label{fig:SafterSecondMilling}}
\hspace{20pt}
\subfloat[Holding WLS fiber into the groove.]
{\includegraphics[width=0.36\textwidth]{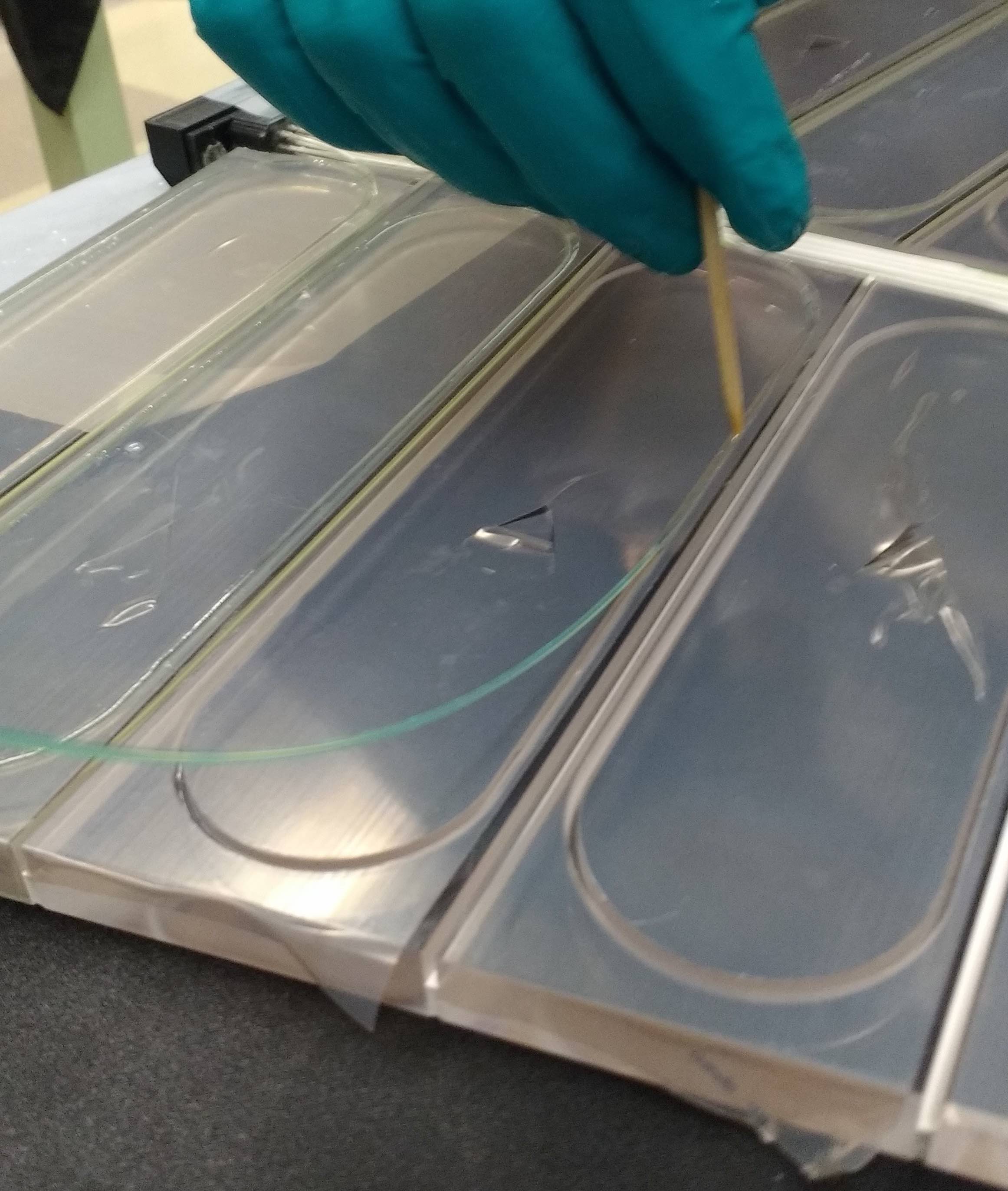}
\label{fig:SholdingDown}}\\
\subfloat[A polished edge near the tip.]
{\includegraphics[width=0.7\textwidth]{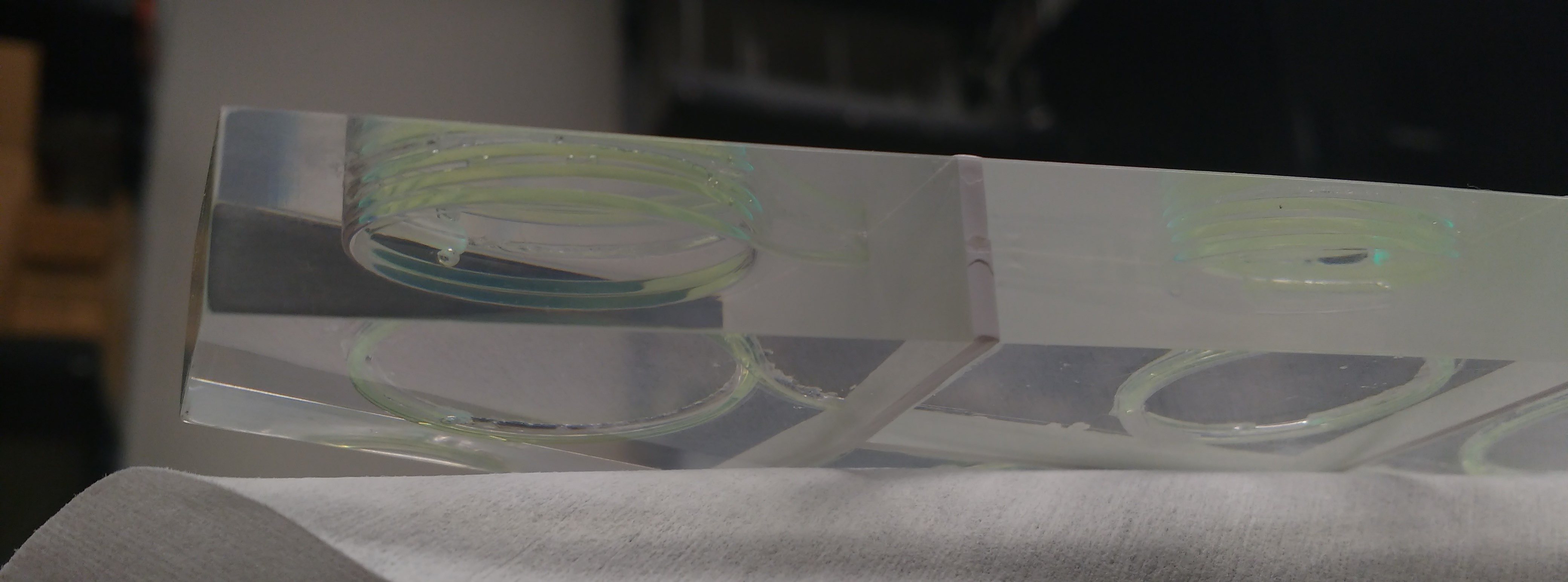}\label{fig:Spolished}}
\caption{
\ref{fig:SedgeTape}: Tape being removed from the edge after
the reflective epoxy in the isolation grooves has dried.  This
is after the first machining stage; later milling and epoxying from the other side will result in complete optical isolation of
adjacent tiles.  Common dust is visible on the face of the
scintillator.
\ref{fig:SafterSecondMilling}: Immediately after the second milling stage.
\ref{fig:SholdingDown}: a WLS fiber is being cemented into
its groove.  PTFE tape is used to protect the face of the
scintillator and isolation grooves.
\ref{fig:Spolished}: The sides have been polished with ${\rm Al_2O_3}$; the WLS fibers for the smaller tiles are visible.
}
\label{fig:ManyScintillatorPhotos}
\end{figure}

In the first machining stage, the isolation groove pattern is milled half-way through
  the 1.2~cm depth of the supersector.
These grooves were filled with diffusively reflective epoxy (see section~\ref{sec:SreflectiveEpoxy}) after first washing
  the supersector thoroughly with water and gentle detergent (with special attention to clearing and wiping
  machining fluid from inside each deep groove) and then taping 
  (Ultra-Low-Friction Tape with Teflon PTFE) its edge
  as a barrier against the epoxy running out of the groove; see figure~\ref{fig:SedgeTape}.
The supersector was held flat with distributed weights to prevent warping
  due to contraction of the drying epoxy over 2--3 days.
  
The supersector is returned to the CNC with its glued side facing the table.
(Before clamping it to the table, it was important to carefully scratch away any
  dried epoxy meniscus that could slightly protrude from the groove and prevent
  a perfectly flat lie.)
The remainder of the isolation grooves, the fiber grooves and ramps, and the 
  central channel are machined in this second step.

After washing off residual machine fluids as before, the next steps are glue-related.
Optically-isolating (reflective) epoxy and optically-coupling epoxy must be
  injected in closely-space
  grooves, and it is very difficult to remove epoxy from a deep groove in
  scintillator.
To avoid contamination, various techniques were improvised which are too
  specialized for general interest.
It is worth noting, however, that the aforementioned PTFE tape was quite
  useful here, due to its strength and low residue, as well as the fact that
  no epoxy sticks to it.
  
The end connector of the WLS fiber bundle (see section~\ref{sec:Sfibers})
  was epoxied to the edge of the supersector, and the WLS fibers were
  glued into tile fiber grooves with Eljen EJ-500 optical cement.
Once mixed, the cement remains usable for roughly 20~min.
(The manufacturer cites a 60-minute working time, but injection by needle
  becomes difficult sooner.)
Mixing tends to introduce small, undesirable bubbles that often do not
  come up quickly.
We found it best to mix the cement in very small batches ($\sim10$~ml),
  and immediately degas it under the light vacuum of a roughing pump for up
  to 30 seconds, before pouring it smoothly into a
  10-ml plastic syringe for injection.
As discussed in section~\ref{sec:SreflectiveEpoxy}, it was
  important to remove any lubricant from the syringe first.
  
The WLS fiber groove was half-filled with cement before the fiber itself
  was slowly laid in, looping three times as mentioned above.
This was a two-person job, with one person gently holding down a portion
  of fiber to keep it from popping out and causing bubbles,
  as shown in figure~\ref{fig:SholdingDown}.
Soft wood or bamboo skewers were used, to avoid damaging the cladding of the fiber.

The cement is left to dry for 36-48 hours, and then the remaining
  isolation grooves and central channel are filled with
  reflective epoxy.
This is similar to filling the isolation grooves from the first
  stage, but the volume of epoxy at this stage is much higher,
  and there is considerably more flow between regions to fill.
The method that worked best was to partially fill the appropriate spaces with
  the white epoxy, allow to dry for 1-2 days (weighted down
  for flatness), and then complete
  the job.
Importantly, a WLS fiber is encased in reflective epoxy from
  the point that it leaves its tile, until it reaches the 
  connector at the outer radius of the supersector.
Near the connector, half of the fibers rise above the face of
  the supersector, so there is a 5-mm ``mound''
  of white epoxy in that region.
  
The sides of the supersector are polished gently with fine sandpaper
  followed by a damp microfiber rag with 1200-grit 5-micron
  ${\rm Al_2O_3}$ lapping grains, until the sides are clear and smooth; see figure~\ref{fig:Spolished}.

Each supersector was finally wrapped in one layer of Tyvek (DuPont 1055B polyethylene fabric 7-mil thick) to
  increase light collection~\cite{Filippov:2000kma} and two
  layers of 10-mil-thick black paper to shield from external light.
(Directing bright light onto the face of a wrapped supersector will, however, generate some current in the SiPMs.
For use in a brightly-lit environment, additional wrapping would be needed.  
In STAR, however, the EPD is shielded from light on one side by its mounting structure;
  the other side faces the dark interior of the experiment.)

Mounting tolerances (discussed below) allot only about 1~mm for wrapping material, so the wrapping had to be tight;
  tight wrapping also prevents optical crosstalk.
Even so, some places (e.g.~near the connector) were regularly prone to light-leaks.
This was discovered when testing the first supersectors, by observing increased current in the SiPMs when flashlights were shone in these areas.
To alleviate this problem, we began to paint those areas with several coats of reflective paint (Eljen EJ-510), prior to wrapping.
Especially near the connector region, extra wrapping and light-shielding was used.

This construction process was relatively straightforward, and 30 supersectors
  were built with very few problems.
In one case, a WLS fiber was broken while wrapping it into the groove of a tile
  in row 2.
Tiles in row 2 are the smallest, and the circular fiber
  groove had a 2-cm diameter, much smaller than
  the 5-cm minimum diameter suggested by the
  manufacturer.
Violating this minimum may lead to breaks or cracks in the fiber; while
  this occurred only once for the EPD, we would not use such low-radius
  bends for these fibers next time.
Additionally, for these S-type fibers, up to 40\% light loss is
  expected due to bending effects; our offline gain measurements confirm this expectation.
Nevertheless, these small tiles-- including the one for which the groove could
  only be partially filled (one-half of one turn) by unbroken WLS fiber-- were
  all usable in practice.

\subsection{Fiber bundles}
\label{sec:Sfibers}
%\red{Here talk about construction of both WLS and clear fiber bundles}
%$\sim1.5$ pages

\noindent \textbf{Optical Fiber Connectors}

All of the connectors for the EPD were 3D printed using black Polylactic Acid (PLA) plastic by an Ultimaker 2+.  This includes the Fiber-to-Fiber Connectors (FFC), Fiber-to-SiPM Connectors (FSC) as well as other necessary pieces such as shields for connectors and a piece that split the clear optical fiber bunch into two sections (c.f. figure~\ref{fig:Soverview}).  After a connector was printed, the holes of the connector had to be drilled out using a drill press as the 3D printer did not have the precision necessary to make holes that would precisely fit the optical fibers.  After this, the surface of the connector was polished as described below, as the outermost surface tended to break away easily, and would scratch the fibers during the fiber polishing process.  Once this step was completed, hex nuts were glued into the connectors so that the connecting screws could be used as the plastic was not robust enough to be threaded.  

\textbf{Design and construction of clear fiber bundles}

Clear optical fiber bundles were constructed in order to connect the EPD wheel itself to the SiPMs (Hamamatsu S13360-1325PE MPPC, $(1.3~mm)^2$, 25~$\mu$m pixels) that were used to read out the signal.  The SiPMs are susceptible to radiation damage, so these were located a distance of 3 meters from the beam pipe, behind the magnet iron.  The Wavelength Shifting fibers (WLS) embedded into the scintillator disks had a short attenuation length ($\lambda > 3.5$~m) and so the connection from the disk to the SiPMs was made with clear optical fibers with a longer attentuation length ($\lambda > 10$~m).  The total length of the optical fiber assemblies was 5.5~m.

The clear optical fibers were Kuraray Clear-PS 1.15-mm-diameter multi-clad fiber.
Each assembly had 31 optical fibers, which matched the distribution of wavelength
  shifting fibers in a given super-sector.  
The fibers were cut, inspected for cracks, 
  and then stretched out into a bundle with no crossing fibers, which was then
  wrapped in Tyvek in order to reduce friction, and slid into the protective UV-Resistant Versilon PVC Opaque Tubing.  
The main section of the tubing was  1/2'' in inner diameter, 3/4'' in outer diameter and 4.8 meters long.  
The fibers were then pushed gently into 31 holes of the fiber-to-fiber (FFC) piece.
The assembly was glued together with Structural Adhesive (Slow-Set Epoxy, 3M 2158 by Scotch-Weld), which included shields that 
  protected the connection between the  FFC and PVC tubing.
The construction of the connector pieces is described in the connector section above.

\begin{figure}[t]
  \includegraphics[width=52mm]{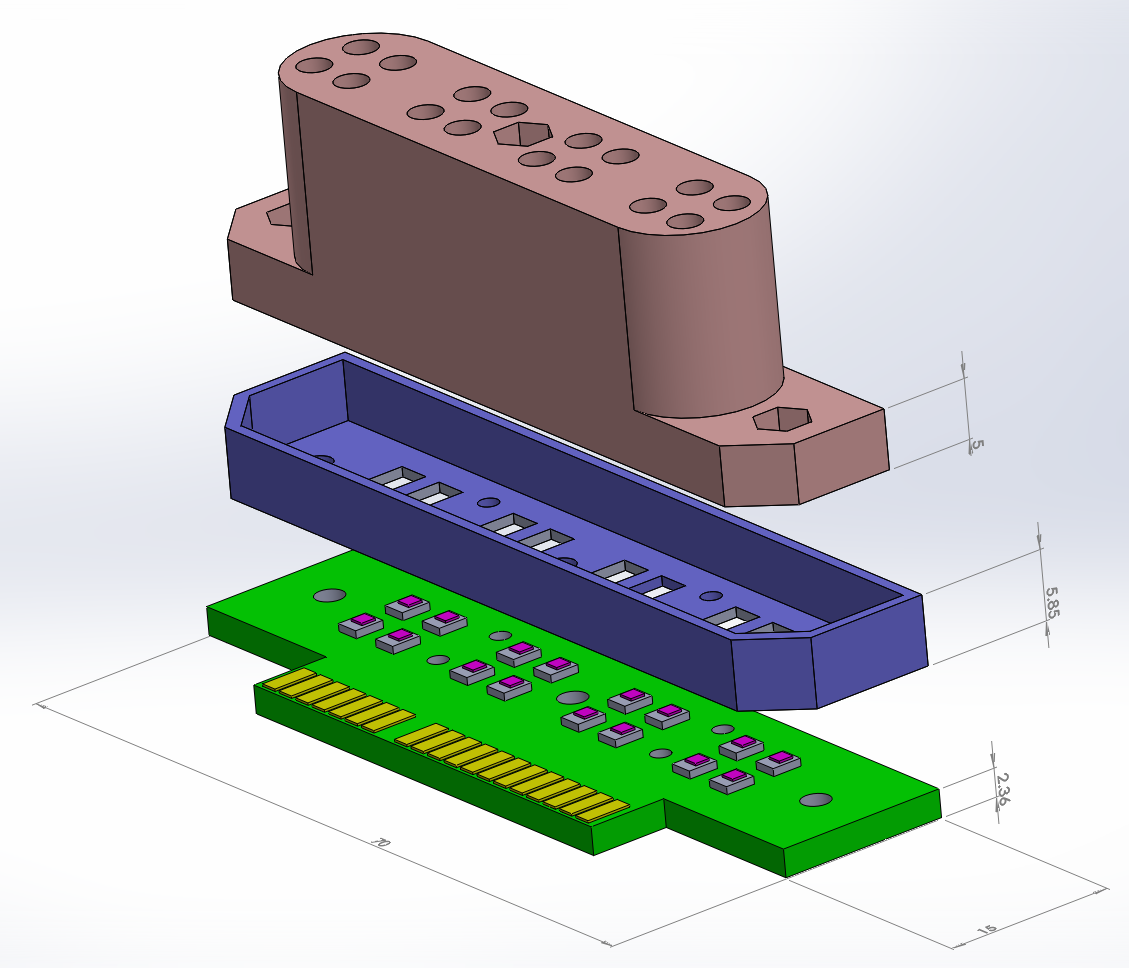}
    \includegraphics[width=44.5mm]{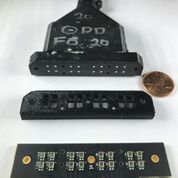}
      \includegraphics[width=59.5mm]{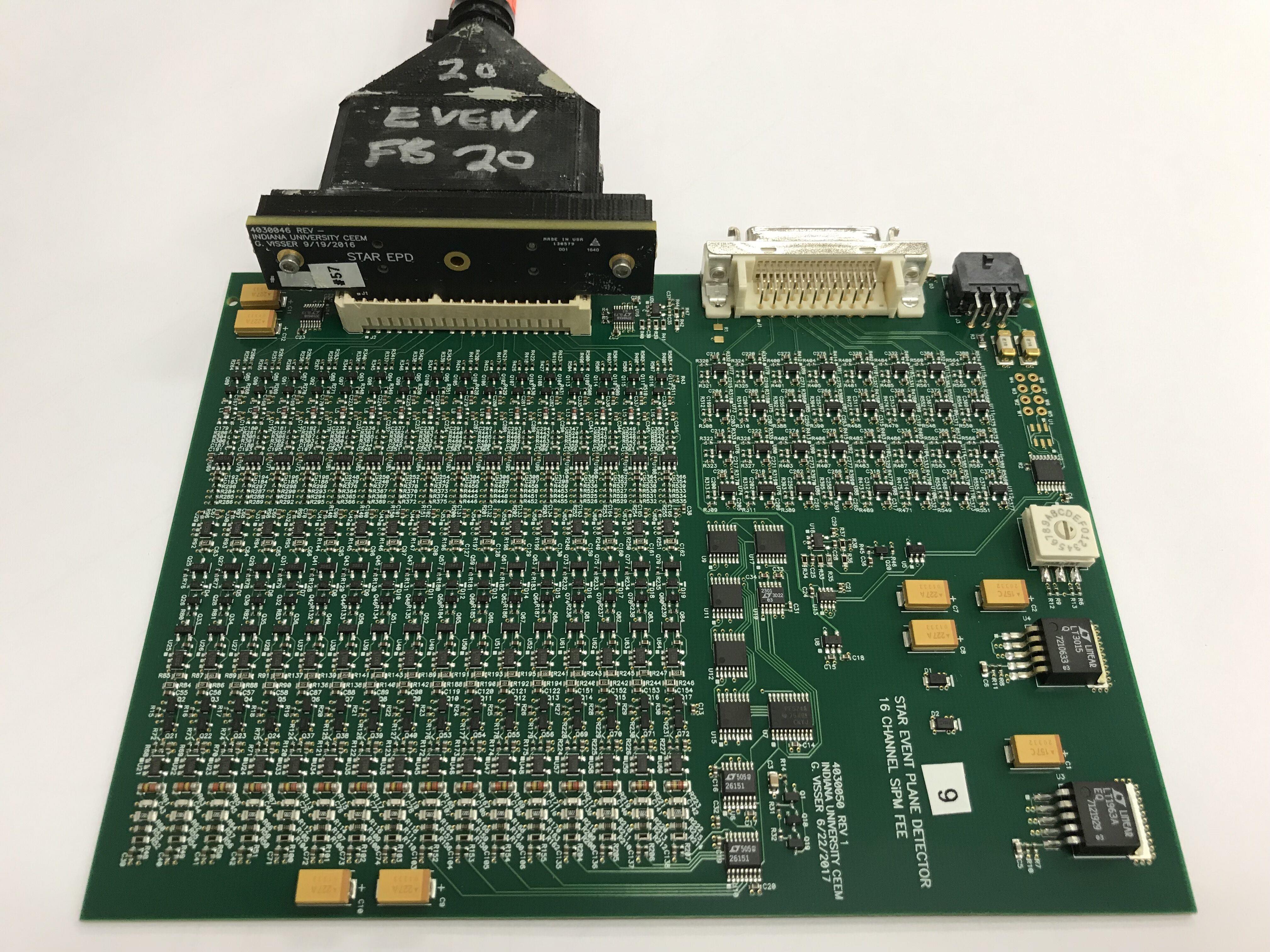}
  
  \caption{On the left is a schematic of the FSC connector, the spacer, and the SiPM board.  In the middle is a picture of the same assembly, with a penny for scale.  The shields, connecting funnel and protective black tubing can also be seen.  On the right is the assembly plugged into the FEE board.}\label{fig:connectors}
\end{figure}

The opposite side of the assembly was split into a group of 15 and 16 fibers, as the read-out was done per half supersector.  A 3D printed splitter was slid down the fibers and connected to the protective tubing via a hose clamp, checking their mapping using an LED light.  Each section of fiber was then encased in a tube of the same PVC material as the main body of the fiber with an inner diameter of 3/8" and an outer diameter of 1/2".  The fibers were funneled into a fiber-to-connector piece, then fibers were carefully routed into two 16 fiber to SiPM connectors (FSC).  This was done by selecting a fiber using an LED light from the FFC end and then sliding it into the appropriate FSC channel.  Once this was completed, each assembly was glued together with the epoxy.  The design of this assembly an can be seen in figure~\ref{fig:connectors}.  After the epoxy dried, the excess fiber was cut and all three ends of the fiber assembly were polished as described below.

\noindent \textbf{Design and construction of WLS fiber bundles}

The WLS fibers were constructed with Kurary Y-11(200) 1~mm-diameter fiber, which we have referred to as WLS fiber. Each fiber in the WLS bundle had to be a different length, in order to correspond to the geometry of the tile that they were glued into.   The length of the fiber within the central channel of the super sector was painted in reflective paint (Eljen EJ-510).  It was found that the best method was to cut, then paint the fibers prior to the construction of the assembly.  Each fiber was fed into the appropriate hole of the FFC fiber connector and then epoxied in.  After the epoxy dried, the excess fiber on the connector end was trimmed and then the fibers were polished.  On the far end the fibers were individually polished and the end was painted in reflective paint.

\noindent \textbf{Polishing}

The optical coupling between the WLS and clear fiber connectors was extremely important, 
  so the fiber assemblies were polished to improve the optical coupling.
All polishing work was done by placing the polishing paper on top of a
  flat glass plate in order to insure that bumps or imperfections in the
  environment would not affect the polishing quality.
First the connector and fiber assemblies were polished with 2500 grit paper.  
Once the largest features on the fibers were removed, we then used diamond lapping sheets, moving down 
  through 30, 6, 3, 1~$\mu$m grit sizes.
At each stage, the fibers were inspected using a microscope.

\noindent \textbf{Testing and Final Assembly}

Each assembly (both clear and WLS) was tested by shining light via an LED through the assembly, and then measuring transmission through the assembly using a photodiode.  Both the single fiber transmission and the average transmission over the entire assembly were examined.

The SiPM boards were designed to be (and remain) attached the clear fiber bundle assembly.  A 3D spacer was designed in order to protect the SiPMs, this and the SiPM board were attached to the FSC ends of the fiber bundles as can be seen in the left and middle pictures in figure~\ref{fig:connectors}.  The SiPM board then would connect to the Front End Electronics (FEE) board via an edge connector, which can be seen in the right of
figure~\ref{fig:connectors}.

%\begin{figure}[t]
 % \includegraphics[width=70mm]{OpticalFiber/MainConnectors.jpg}
  
  %\caption{The WLS connector is on the left and the clear optical fiber is on the right in this assembly.  The metal alignment pins can be seen separating the two connectors.  The FTC and shields can be seen on the right optical connector, as well as the black protective tubing. }\label{fig:connectors}
%\end{figure}

\subsection{Reflective Epoxy}
\label{sec:SreflectiveEpoxy}

The success of this detector relies on the reflective consistency and strength of
  the isolating reflective epoxy.
The procedure to produce, store, and use the epoxy was based largely on   
  the detailed discussion by Olsson, {\it et al}~\cite{Olsson:1995qa}.
However, that manuscript is difficult to find in print, and was accompanied
  by passed-down clarification.
Therefore, we hope it is useful to archive our procedure here.

We used D.E.R. 332 epoxy resin with Jeffamine D230 curing agent and ${\rm TiO_2}$
  powder (44 micron).
For the mixing, we used a small ball mill with a thick glass 500~ml jar.

Insert 100~mg of resin into the jar together with
  8--10 mixing balls ($\tfrac{1}{2}^{\prime\prime}$ diameter); we found it important to use stainless steel balls to avoid discoloration of the white epoxy.
Warm in a lab oven at $62^\circ$C for 3 hours, then remove and add 50~g
  of ${\rm TiO_2}$ powder.
  
Tightly seal the jar and spin (our mill spun at 40~rpm, 
  though Olsson~\cite{Olsson:1995qa} suggests 20~rpm) on the ball mill for 3 hours.
Remove from the mill, loosen the lid, and place jar back in the $62^\circ$C oven
  for 2 hours.
Tighten the lid and repeat mixing on the ball mill for 3 hours,
  then once more loosen the lid and place in the oven for 2 hours.

Remove the jar from the oven, add 32~g of Jeffamine, and mix in the ball mill
  for 2--3 minutes.
Portion the epoxy directly into syringes.
The filled syringes should be used immediately or flash-frozen in
  liquid nitrogen and transferred to a freezer (about $\-20^\circ$C)
  for later use.
These syringes may be re-warmed in the $62^\circ$C oven ($\sim15$ minutes or until
  the epoxy flows) for up to several weeks afterward.
  
We used inexpensive 10~ml plastic syringes with a silicone rubber piston stop.
We found that some batches of the syringes had a small amount of lubricant in the plunger which could slightly discolor the epoxy or optical cement.
This is easily avoided by simply removing the plunger and wiping the stop before returning it and filling.

The reflecting epoxy was injected by hand through
  stainless steel dispensing needles ($1^{\prime\prime}$ 21-gauge) coupled to the syringe by luer lock.
Great care was taken to avoid air bubbles in the isolation grooves.
Ten ml of warm epoxy flows smoothly through the needles; warming mid-way through
  use was sometimes helpful.
  
\begin{figure}[t!]
\subfloat[A sextant support structure holds two supersectors.]
{\includegraphics[width=0.45\textwidth]{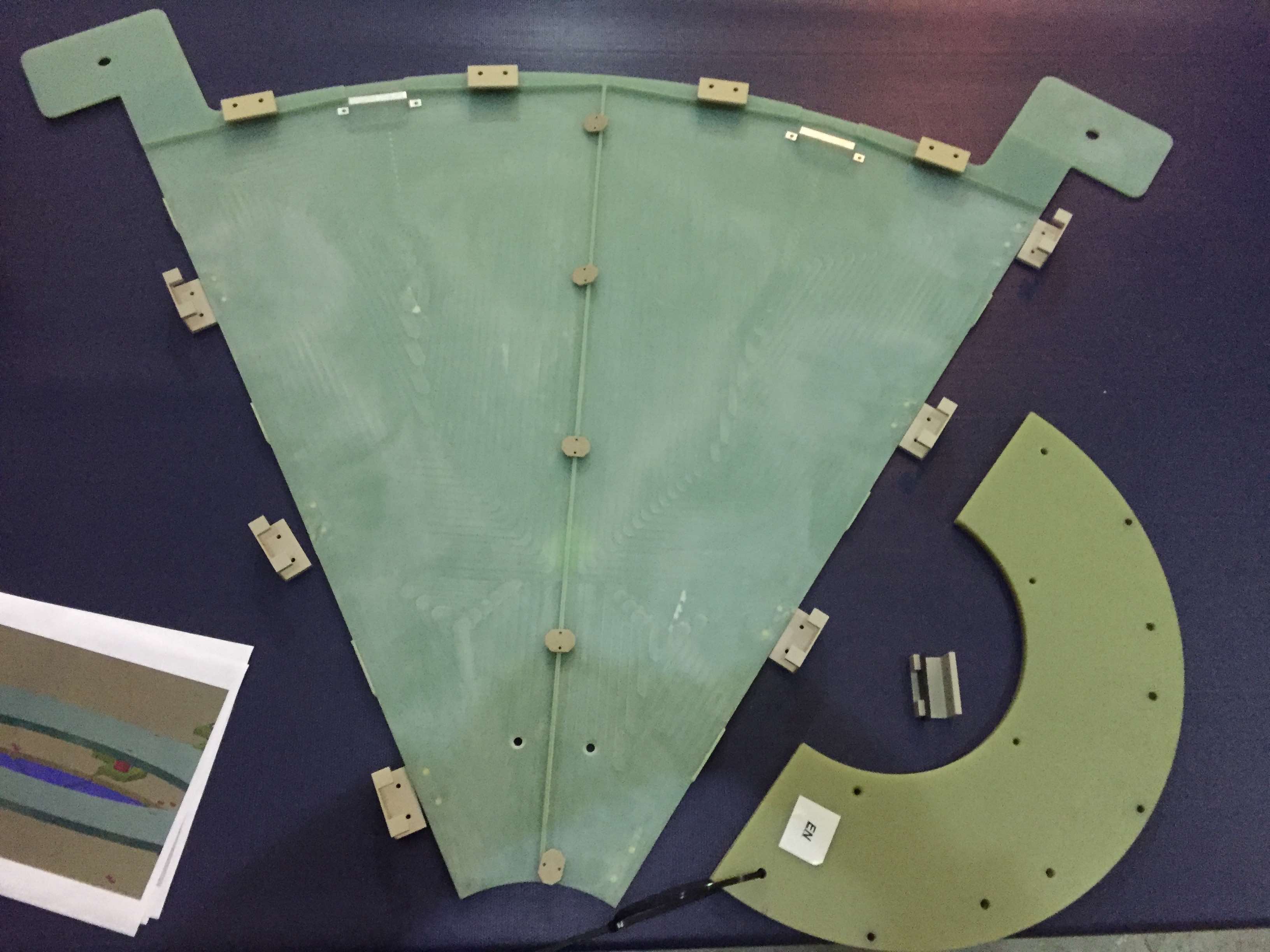}\label{fig:SstrongbackAlone}}\hspace{1mm}
\subfloat[Supersectors mounted in strongback.]
{\includegraphics[width=0.45\textwidth]{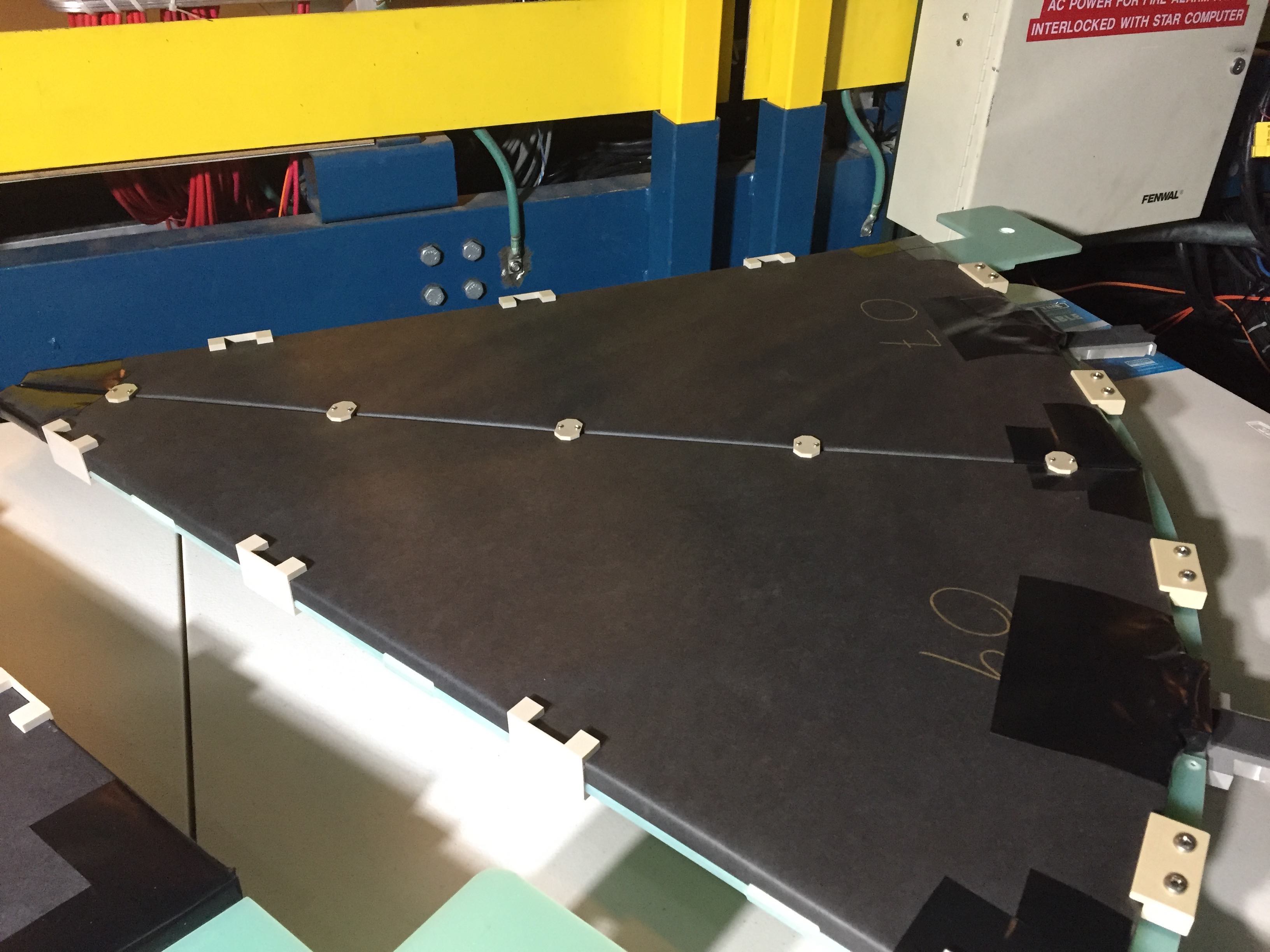}\label{fig:SsupersectorsInStrongback}}\\
\subfloat[Mounted on the West side of STAR.]
{\includegraphics[width=0.45\textwidth]{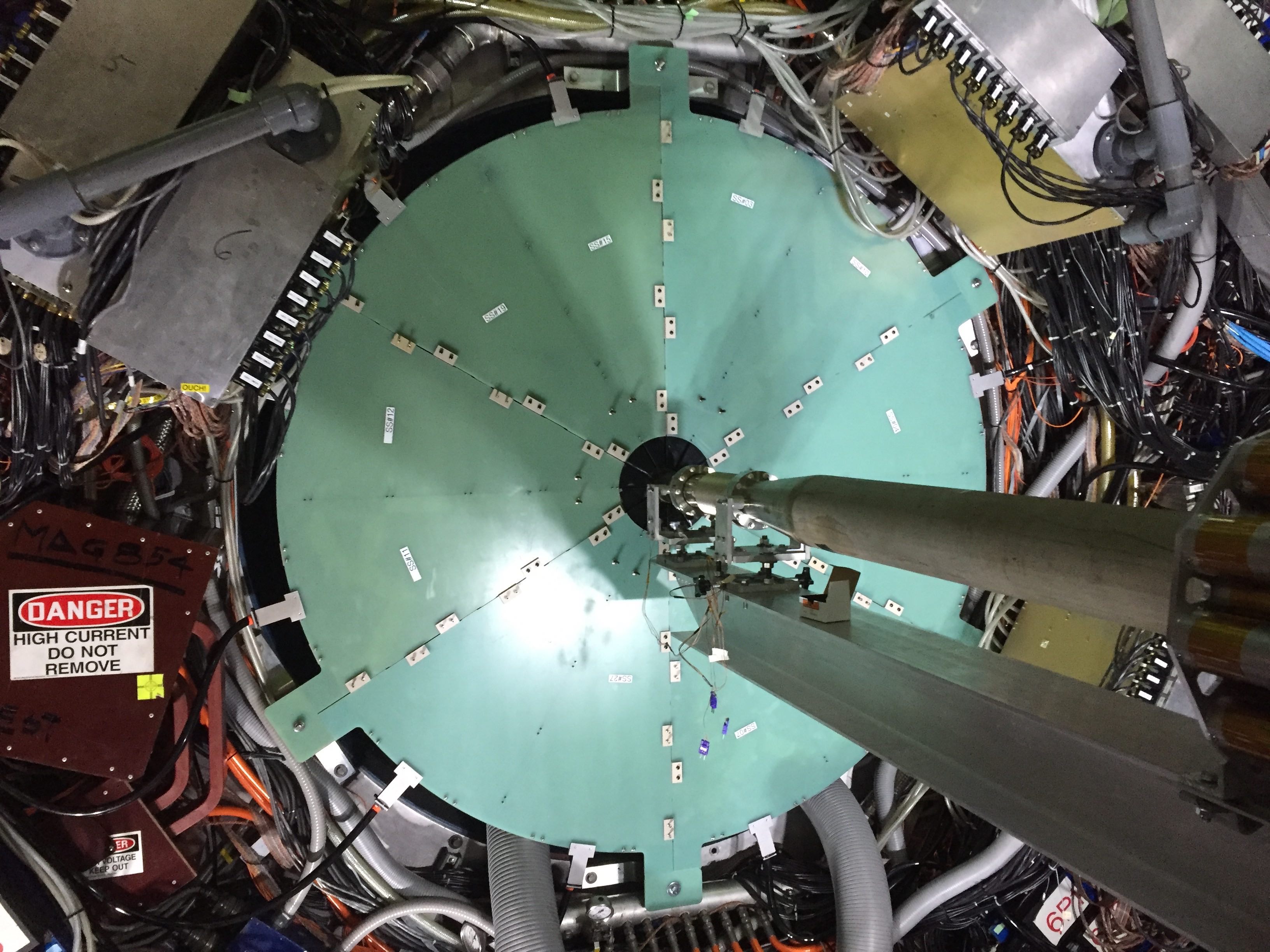}\label{fig:SmountedOnWestSide}}\hspace{1mm}
\subfloat[Zoom on inner tips near beampipe.]
{\includegraphics[width=0.45\textwidth]{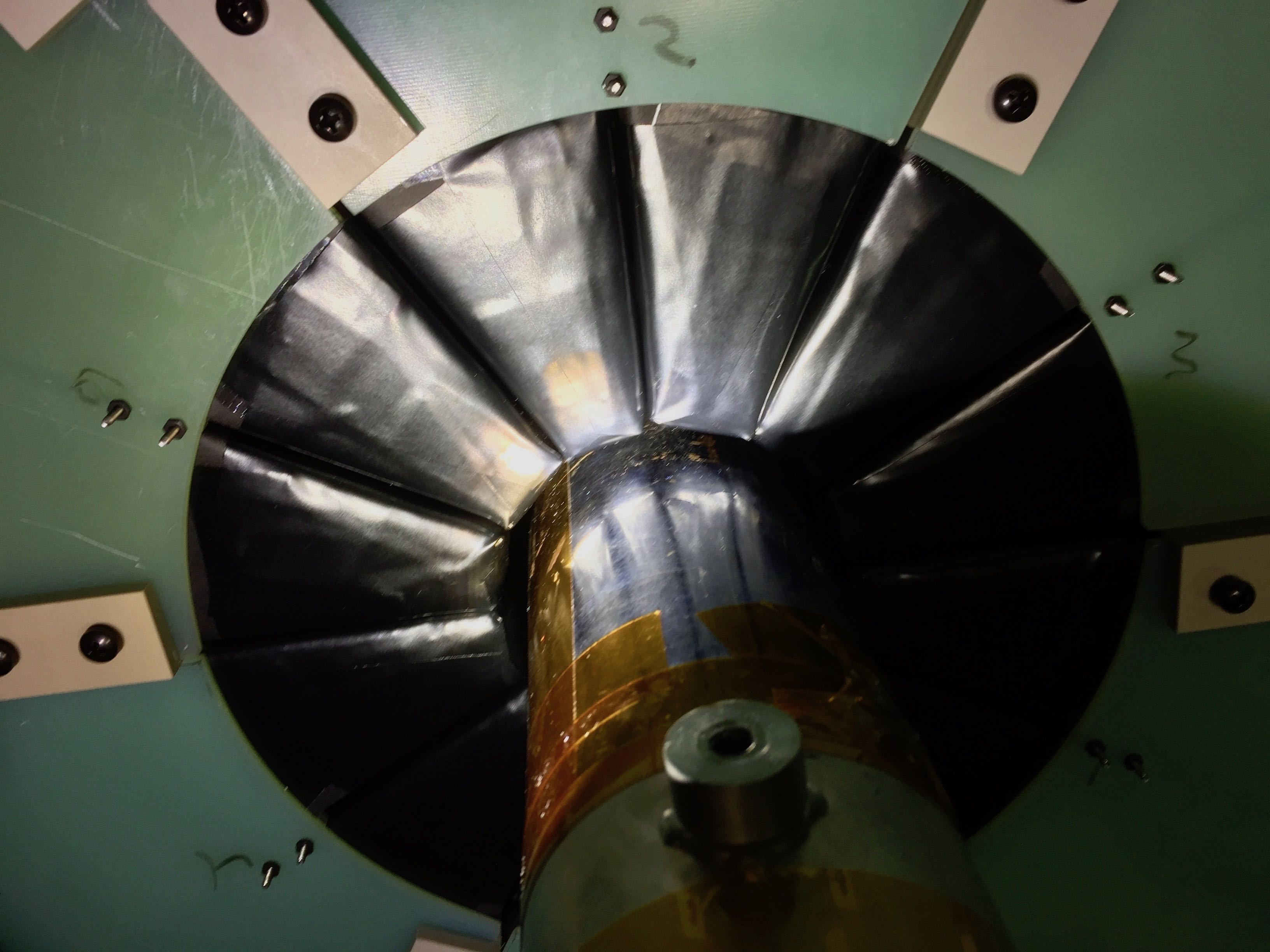}\label{fig:nearBeampipeZoom}}
\caption{The EPD mounted onto the experiment.}
\label{fig:SmountedInStar}
\end{figure}

\subsection{Mounting structure and material budget}
Mounting the EPD in the STAR experiment is nontrivial.
The experiment is fairly bristling with detector systems, and there is limited space outside of the
nominal volume of the wheels themselves; see figure~\ref{fig:SmountedOnWestSide}.
Mounting infrastructure must allow close detector packing.
Consideration must be given to installation procedures on manlifts in restricted areas near the beampipe;
  awkward reaching must be minimized and the mounting position must be reproducible.
Furthermore, it is important to keep the material budget
down, as there are detectors (and future upgrade
detectors) ``downstream'' of the EPD.

The mounting scheme is seen in figure~\ref{fig:SmountedInStar}.
Each support assembly 
  consists of six adjacent sextants milled from 3/8 inch-thick fiberglass-reinforced epoxy laminate sheets, FR-4. 
This material is stiff enough to 
  support two  supersectors in each sextant, and the resin binder in FR-4 is especially flame resistant.  
Supersectors are placed in recesses
  milled into the front of the 
  support structure and held in place with C-shaped brackets along the edges that use screws to pin the supersector in position and 
  tabs along the center spoke of each sextant.
The design is secure but introduces negligible mass
  between the EPD and the collision vertex.

Each sextant mounts on two of six mounting studs on the
  STAR magnet pole tip, as seen in figure~\ref{fig:SmountedOnWestSide}.
The inner tips of the EPD come within 5~mm of the beampipe, as seen in figure~\ref{fig:nearBeampipeZoom}), and to minimize material in front of detectors further downstream,
  the support structure ends about 10~cm from the 
  supersector tip.

\noindent{\bf Contribution to STAR's mass budget}\\
The scintillator (polyvinyltoluene) has a radiation length of 42.54~cm~\cite{PVT} and the FR-4 of the support structure has a radiation length of 
  19.4~cm~\cite{G10}. The scintillator is 1.2~cm thick. The depression of the support structure which holds the scintillator is 0.45~cm. Outside of this
  depression the support structure is 0.95~cm thick. Thus particles passing through the scintillator pass through 0.051 radiation lengths and particles 
  passing outside of the scintillator pass through 0.049 radiation lengths.
  
%%%%%%%%%%%%%% Section 3 - Tests and performance

\section{Tests and Performance}
\label{sec:TestsPerformance}

Here, we describe bench tests performed in the lab, in parallel with construction,
  and performance of the system as deployed in the STAR experiment.

\subsection{Bench tests of completed supersectors}

Two bench tests were performed on each tile. 

\noindent
{\bf Cosmic ray test stand}\\
Four supersectors were aligned in a compact vertical stack, between
  pairs of paddle scintillators read out by photomultiplier
  tubes and fed into a leading-edge discriminator.
The noise rate 
  in each paddle detector was on order 2~kHz, and the pulse width
  on order 10~ns, so a coincidence between them is a clean
  trigger for cosmic rays.
Using external trigger detectors with areas on order 200~${\rm cm^2}$, we were able to simultaneously test about 20--28 tiles (5--7 tiles from each of the four supersectors)
in a $\sim10$-hour-long run.

\begin{figure}[t]
  \includegraphics[width=70mm]{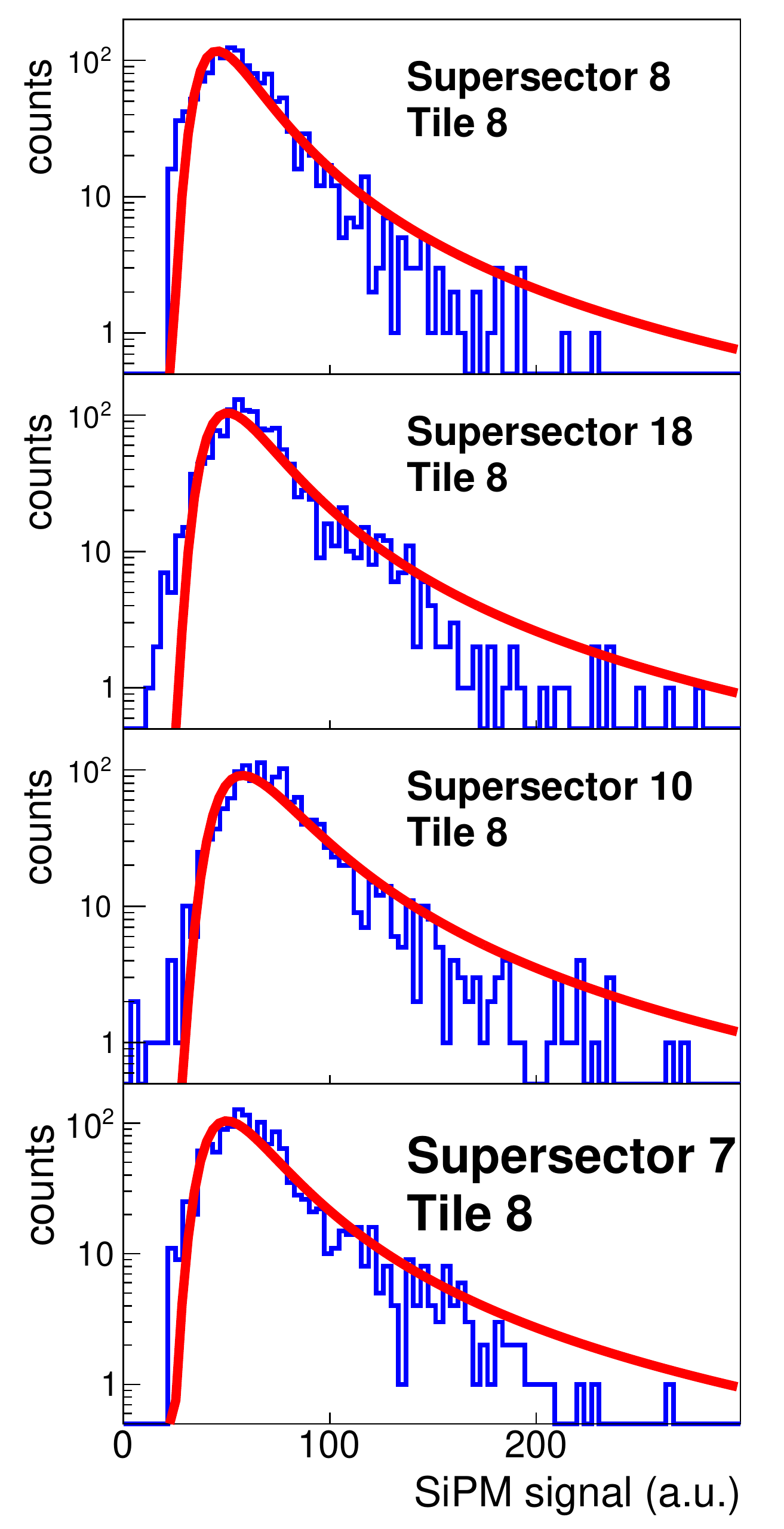}
  \caption{
  Spectra (in ADC units) from the cosmic ray test stand.
  A coincidence between large external scintillator  paddles
  triggered the CAMAC acquisition system.  Above-threshold signals in the top (here, supersector 8) and bottom (supersector 7) tiles indicated a roughly vertical cosmic ray through tile 8.
  Landau fits are drawn for reference.
  \label{fig:ScosmicTestStand}
}
\end{figure}

Spectra for four vertically-aligned tiles obtained in the test stand are shown in figure~\ref{fig:ScosmicTestStand}. 
Spectra were measured with LeCroy 2249A ADCs using a 160-ns gate (more than sufficient to fully integrate the $\sim50$-ns-FWHM signal).
The trigger detectors covered a range of tiles, so for a given stack of tiles, a signal above threshold in the top and bottom tiles (from supersectors 8 and 7, in this figure) was identified as a ``vertical ray event''.
Spectra from vertical rays were fitted with roughly the expected shape (Landau 
  distribution) for reference.
  
The test stand system was relatively noisy and statistics were limited.
These tests were intended mostly to flag gross problems and identified about 4 or 5 problem tiles, usually exhibiting low light efficiency;
  only two of the 930 tiles examined were deemed unacceptable for use.
  
Higher-statistics spectra are discussed in section~\ref{sec:Sperformance}

\noindent
{\bf Detector uniformity and isolation scan}\\
The EPD is, to first order, a ``hit detector,'' and the fine segmentation
  means that regions of scintillator on order 20~${\rm cm^2}$ (tiles) are
  separately gain-equalized.
Therefore, unlike scintillator sheets in an electromagnetic
  calorimeter~\cite{Allgower:2002zy} for example, uniformity of response
  as a function of position is not a big concern.
Nevertheless, we performed extensive tests to measure the position dependence
  of the EPD response using a narrow beam of $\beta^-$ radiation from a 200-$\mu{\rm Ci}$ $^{90}{\rm Sr}$ source.

\begin{figure}[p]
\subfloat[Response of tiles 5,6,7 and 9 in one supersector, as a function of source position.  Colors indicate background-subtracted current in $\mu$A.]
{\includegraphics[width=0.9\textwidth]{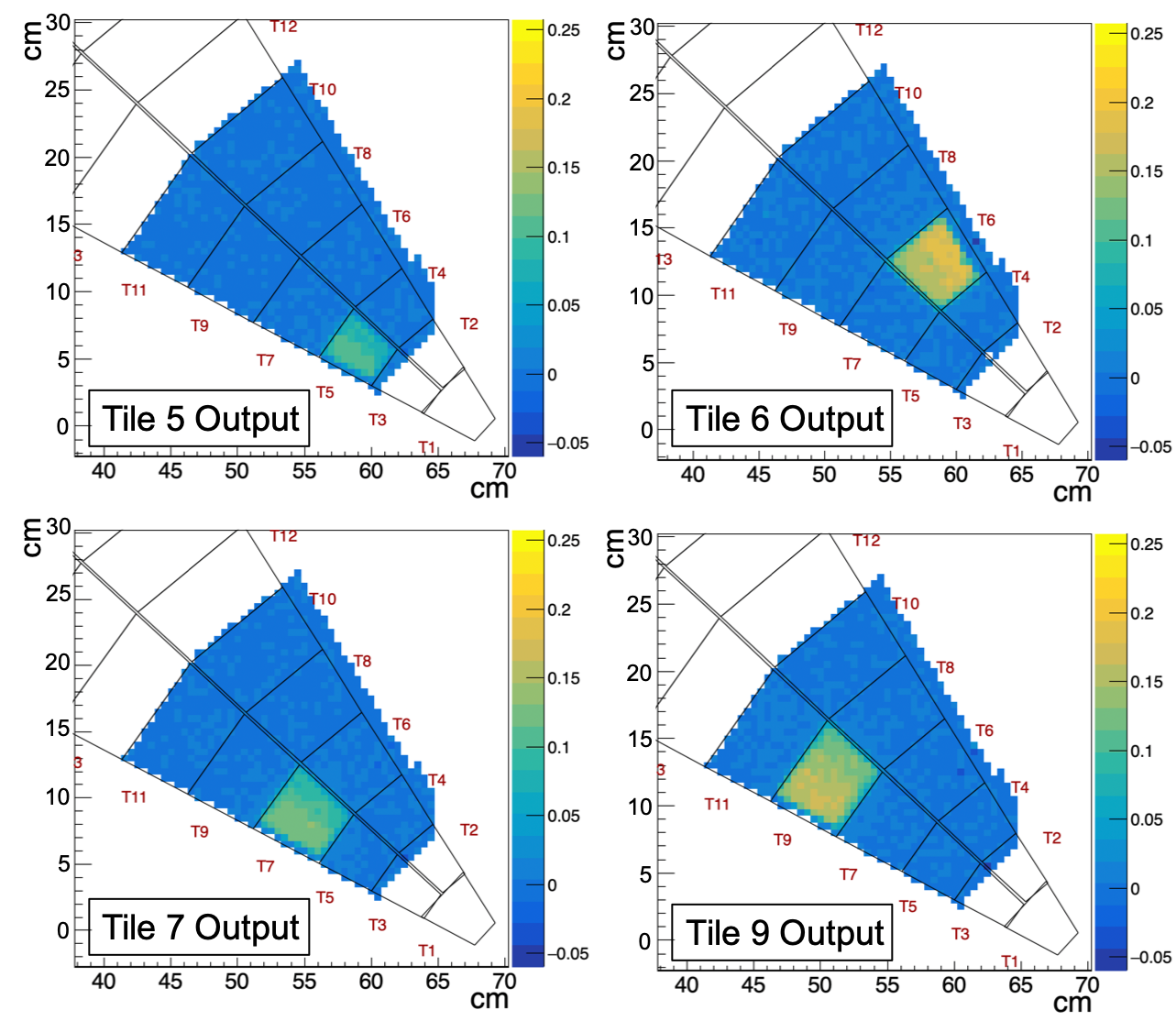}\label{fig:SsourceScan2D}}\\
\subfloat[Signals in radially adjacent tiles, as the $^{90}{\rm  Sr}$ source scans in the radial direction.]{\includegraphics[width=0.48\textwidth]{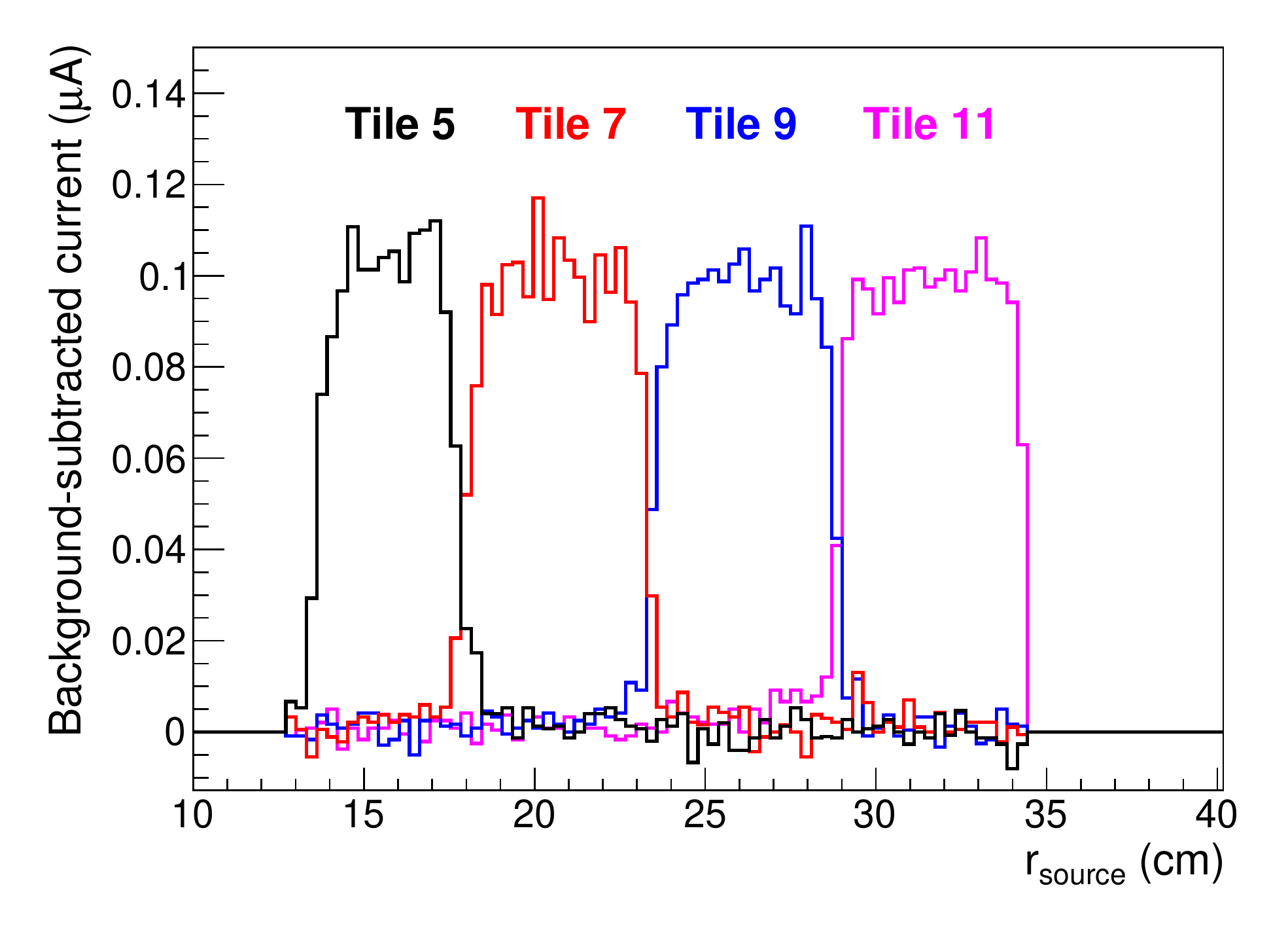}\label{fig:SsourceScanRadial}}\hspace{0.1cm}
\subfloat[Signals from tiles 8 and 9, as the $^{90}{\rm  Sr}$ source is moved in a path perpendicular to the central channel.]
{\includegraphics[width=0.48\textwidth]{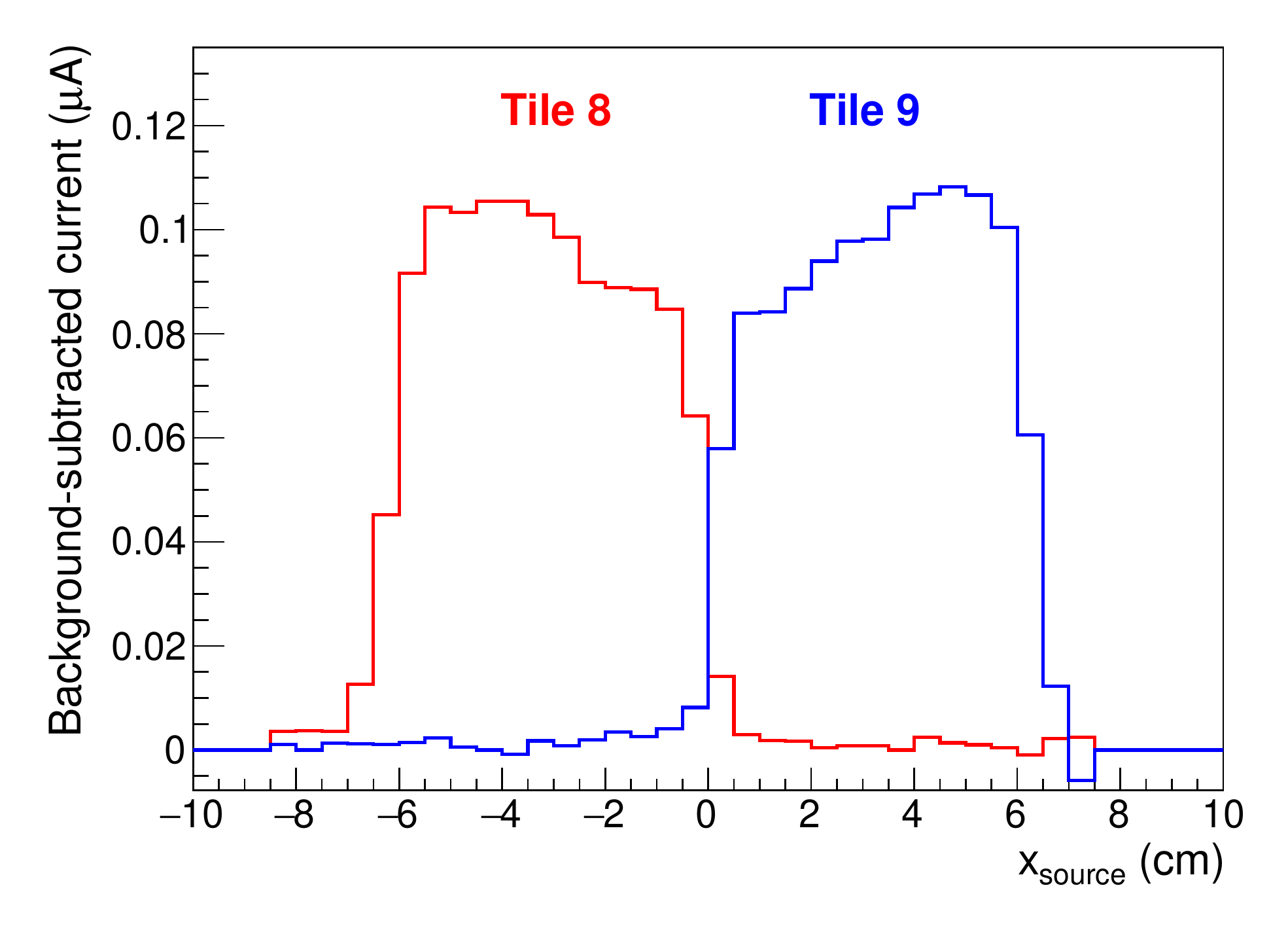}\label{fig:SsourceScanPerp}}
\caption{
Detector response (SiPM current above background) as a function of position of a $^{90}{\rm Sr}$ source.  See text for details.
}
\label{fig:SsourceScan}
\end{figure}

The source was mounted on the platform of a $x-y$
  translation table driven by computer-controlled stepper motors.
The platform could travel 80~cm in each direction.
A supersector was held above the table with its face about 3~cm from the source,
  and the current of the biased SiPM was read out.
  
An automated program drove the source to a specified spot beneath the supersector,
  read out the current of all 31 SiPMs, 10 times, and recorded the average.
It then drove the source to a position away from the supersector, read out all
  31 channels 10 times, and recorded the average; in this way, the baseline
  current (dominated by small light leakage from ambient room lighting as well as small dark current in the SiPM itself), 
  which could in principle vary with time, could be subtracted.
This procedure was repeated for a grid of points on the surface of the supersector.
Grid spacing varied from about 5~mm to about 2~cm, and full scans ran
  for about 12 hours.
We scanned most supersectors, at least coarsely.

Representative results are shown in figure~\ref{fig:SsourceScan}.
Each panel in figure~\ref{fig:SsourceScan2D} represents the increase in (i.e. background-subtracted, as discussed above) current of a single SiPM
as a function of position of the $^{90}{\rm Sr}$ source.
Uniformity of response over the face of a tile was very good in the radial direction and typically better than 25\% in the direction perpendicular to radial, perfectly acceptable for our purposes.

The degree of optical isolation from light induced in a neighboring tile was excellent.
Figure~\ref{fig:SsourceScanRadial} shows the response of four adjacent tiles, as the source is moved along a line in the radial direction.  The transition from one tile to the next takes place within about 1~cm, which is also about the size of the spot
illuminated by the $^{90}{\rm Sr}$ source.
Therefore, the level of true light signal ``crosstalk'' across these tiles is negligible.
In figure~\ref{fig:SsourceScanPerp}, the source is moved across the face of tile~8 into tile~9, along a path perpendicular to their mutual border.  Again, the isolation is excellent; any optical crosstalk is below the sensitivity of this technique.
This figure also highlights the $\sim20\%$ non-uniformity of response in this direction, typical of most tiles.

Crosstalk on the level of $\sim5\%$ was observed between two tiles in two supersectors; this is believed to be the result
  of a drop of optical cement coupling a small length of WLS fiber exiting one tile onto the edge of a neighbor.
This level of crosstalk has negligible impact on EP determination, as the signals are still associated with similar azimuthal angle.

\subsection{Performance in STAR}
\label{sec:Sperformance}

The full Event Plane Detector was first
  installed and run in STAR during the 2018 RHIC campaign,
  with all 744 tiles performing well throughout the
 campaign.
It was also installed for the 2019 campaign, in which it
  functioned equally well.

\begin{figure}[t]
\subfloat[Uncalibrated spectra from ring 4.]
{\includegraphics[width=0.5\textwidth]{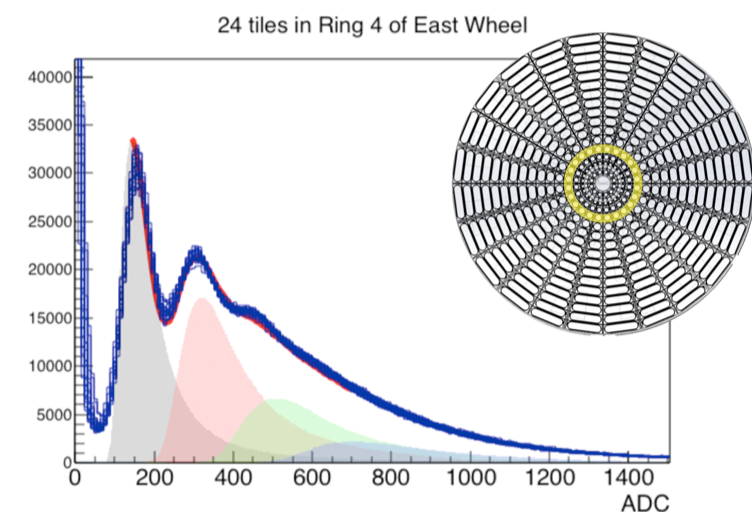}\label{fig:EastWheelRing4}}\hfill
\subfloat[Uncalibrated spectra from ring 14.]
{\includegraphics[width=0.5\textwidth]{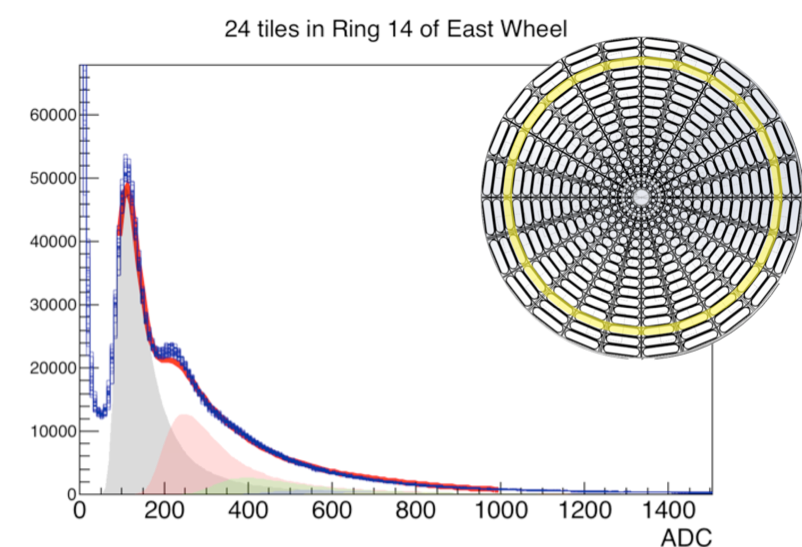}\label{fig:EastWheelRing14}}
\caption{
Twenty-four uncalibrated spectra superimposed (thin blue lines) from the 24 tiles of rings 4 (\ref{fig:EastWheelRing4}) and 14 (\ref{fig:EastWheelRing14})
of the east wheel of the EPD; the individual spectra are nearly indistinguishable.
The spectra have been scaled along neither the vertical nor the horizontal axis.
Shaded regions indicate Landau energy loss calculations corresponding to one, two, three, or four minimum-ionizing particles simultaneously
depositing energy in the tile.
A red curve, mostly hidden behind the spectra themselves, indicates the sum of these contributions.
}
\label{fig:GainResponseUniformity}
\end{figure}

\noindent \textbf{Energy loss spectra and uniformity}

The silicon photomultipliers operate with a nominal bias voltage of 58 V; gain at nominal bias is $\sim7\times10^{5}$
However, by changing the bias by $\sim\pm1$~V, the effective gain may be adjusted within about a factor of
four, without losing linearity of response.
After traveling a short distance along the SiPM card and an edge connector which plugged into a custom amplifier/bias card (see figure~\ref{fig:connectors})
the SiPM signal was further amplified by a low-noise amplifier that preserved the $\sim10$-ns risetime.
The signal was digitized by ADCs with a 75-ns integration time.
We individually adjusted the bias voltage of each SiPM at the start of
the campaign, so that the
  digitized energy loss peak for all tiles is found at the same
  ADC (analog-to-digital converter) value.
Further fine time-dependent adjustments can be performed in offline analysis, but there was little need, as the hardware gains were quite
  stable over the several-month-long campaign; this reflects the stability of the regulated bias supplies on the custom amplifier/bias cards.

Figure~\ref{fig:GainResponseUniformity} illustrates the excellent uniformity of response of the EPD.
Panel~\ref{fig:EastWheelRing4} shows the raw ADC spectra from 24 tiles in ring 4 (c.f. table~\ref{table:EPD_Segmentation}).
While no scaling of yields or gain adjustment (beyond initial bias voltage adjustments, discussed above) were performed, the spectra are nearly identical.
The shaded regions represent calculations of the spectra when one (gray), two (red), three (green) or four (blue) particles pass through a tile simultaneously.
The 1-, 2-, \ldots N-particle distributions overlap considerably, so that for a given collision, the number of particles
  passing through a tile may only be probabilistically determined.

The amount of this overlap is driven by the width of the shaded regions.
The one-particle contribution is a simple Landau distribution whose width is about 14\%--20\% of the most probable value
(MPV).\footnote{It is important to note that the calculated multiple-particle contributions are simply convolutions of the one-particle Landau
  distribution-- their peak positions and widths are not independent parameters.}
This is consistent with irreducible physical energy loss fluctuations expected for a scintillator of this 
  thickness~\cite{Jones:1968xx,Paul:1971xxx}.
This indicates that broadening due to finite photoelectron statistics in the SiPM is not significant; our light-collection
  efficiency was more than adequate for our purposes.
Landau fluctuations also dominate electronics noise effects.
On the bench, the front-end amplifiers were verified~\cite{EpdFeeToBePublished} to have single-photoelectron resolution, and the overall effective noise
in the STAR environment was equivalent to about 2 photoelectrons.

\begin{figure}[t]
  \includegraphics[width=0.5\textwidth]{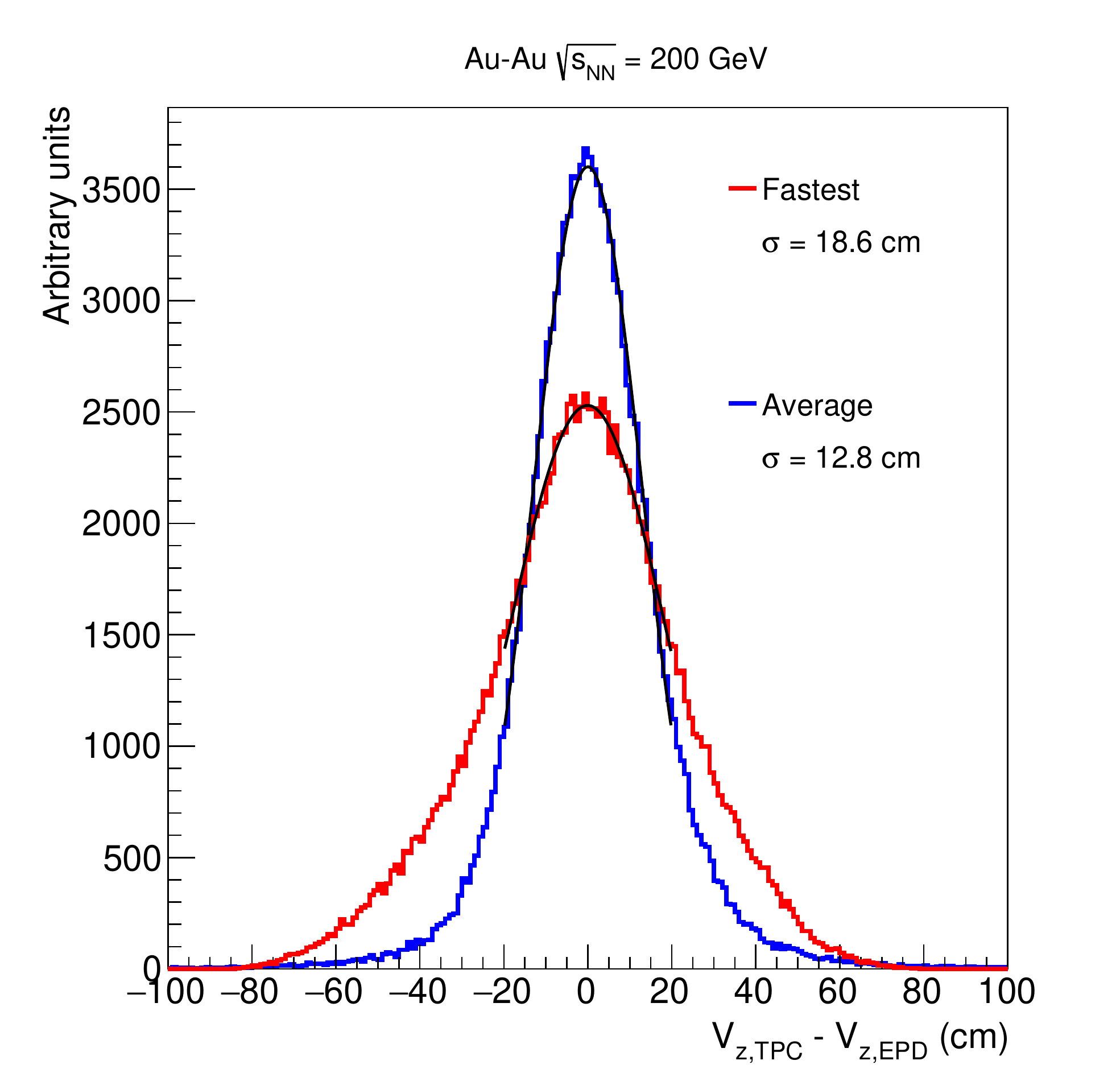}
\caption{\label{fig:timing}
The difference between the primary vertex position as calculated by the TPC and the EPD.
For the red (blue) distribution, the earliest tile time (average of tile times) is used
  to calculate $V_{z,{\rm EPD}}$.
Gaussian fits to the peaks are used to estimate the position resolution from the EPD system.}
\end{figure}

\noindent \textbf{Timing and vertex position resolution}

Depending on the beam properties, ion collisions may occur over a wide range of positions along the beam line within
  the 4-meter-long STAR experiment.
The EPD was used as a trigger detector to quickly identify those collisions which occur close to the center of STAR.
The primary vertex position was estimated as $V_{z,{\rm EPD}} \equiv \left(T_{\rm E}-T_{\rm W}\right)\cdot c$,
  where $c$ is the speed of light.
The time $T_{\rm E}$ ($T_{\rm W}$) is calculated using the slew-corrected times at which tiles in the inner five rings of
  the east (west) wheel fire leading-edge discriminators.

Through multiple-particle tracking in offline analysis, the STAR TPC~\cite{Anderson:2003ur} can determine the position of the collision vertex with
  sub-millimeter precision.
This position is compared with the fast estimate from the EPD in figure~\ref{fig:timing},
  for Au+Au collisions at a collision energy of 200~GeV.
When only the earliest tile on the east and west wheels are used to calculate $T_{\rm E}$ and $T_{\rm W}$,
  the EPD primary vertex resolution is 18.6~cm.
This indicates
  single-tile timing resolution of about $\left(18.6~{\rm cm}\right)/\left(\sqrt{2}c\right)\approx0.44~{\rm ns}$.

When $T_{\rm E,W}$ is the average of all tiles, the primary vertex position resolution improves to 12.8~cm,
  though this will depend on total particle multiplicity.

\noindent \textbf{Event plane resolution}

As discussed in section~\ref{sec:Sintroduction}, the EPD was primarily built to determine the event planes of each collision.
Therefore, without delving into the physics or analysis details, it is appropriate to conclude with an EP-related performance measure.
The so-called second-order event plane is the plane in which the majority of particles are emitted.  For particles emitted to the
  west ($p_z>0$), the azimuthal angle of this plane is given by
  \begin{equation}
    \label{eq:EP2}
      \Psi_{2,{\rm West}} \equiv \frac{1}{2}\tan^{-1}\left(\frac{\sum\limits_{i=1}^{372}w_i\cdot\sin(2\phi_i)}{\sum\limits_{i=1}^{372}w_i\cdot\cos(2\phi_i)}\right) ,
  \end{equation}
where the sum runs over all tiles on the west wheel, and $\phi_i$ is the azimuthal angle of the tile.
In the simplest analysis, the ADC value for a given tile may be used as its weight, $w_i$, though alternate weighting schemes are often employed.

\begin{figure}[t]
  \includegraphics[width=0.6\textwidth]{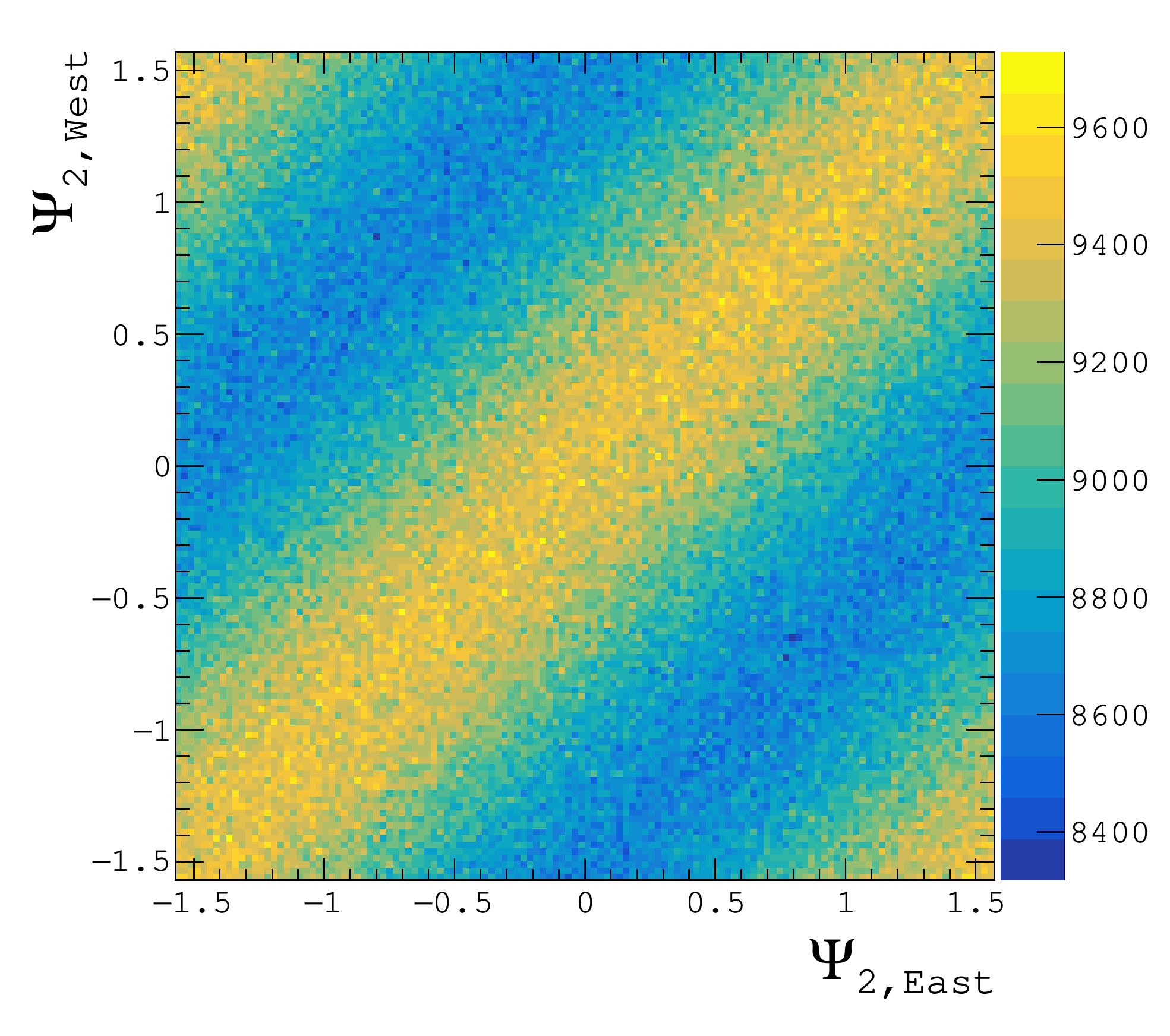}
\caption{
The event-plane angles measured in high-energy nuclear collisions with the east and west EPD wheels show a clear correlation.  See text for details.
\label{fig:EastWestEP2}
}
\end{figure}

Figure~\ref{fig:EastWestEP2} shows the correlation between the west EP angle defined in equation~\ref{eq:EP2} and the east EP angle defined similarly.
There is clearly a strong correlation between the momentum anisotropy in regions separated by many units of pseudorapidity, indicating a ``global'' anisotropy in the initial collision geometry.
The strength of the correlation is a measure of the resolution $R_{EP}$ with which the ``true'' event plane is measured by the detector~\cite{Poskanzer:1998yz}.
The EPD was built to replace STAR's original event-plane detector in this $\eta$ region, the Beam-Beam Counter (BBC).
Depending on the beam energy, event plane type, and other details, the EPD delivers $\sim1.5-2.5$ the event plane resolution of the BBC.
The statistical uncertainty of an EP-dependent physics observable $X$ typically scales as $\delta X\sim N^{-1/2}\cdot R_{\rm EP}^{-1}$, where $N$ represents a statistically relevant number (number of events, number of particles, etc).
Hence, for EP-dependent studies, a two-fold increase in EP resolution is equivalent to a four-fold increase in event statistics.

\section{Summary}
\label{sec:Sconclusion}

A new upgrade detector has been
  constructed and deployed in the STAR experiment at RHIC.
The Event Plane Detector provides STAR with significantly
  improved resolution for measuring the initial geometry of a heavy ion collision, a crucial step in decoding the complex physics of the system's evolution. 

The design and construction of the EPD has been discussed in detail.
The construction was carried out in university labs and machine shop by teams including
  many students and postdocs, providing valuable training in a field sometimes dominated
  by computer-based data analysis.
Cosmic-ray tests of nearly 1000 tiles indicated very few problems with light collection,
  and scanning at the centimeter scale with a radioactive source indicates excellent
  optical isolation of tiles and good uniformity of response.

The EPD has performed solidly in two RHIC campaigns thus far, with all channels functional.
The single- and multiple-hit resolution is as expected for a scintillator of this thickness,
  and the uniformity of response is excellent.
In addition to quantitative improvements in event-plane-related measurements, the
  improvement in event plane resolution may have qualitative implications.
Measurement of a subtle effect that would have previously required a twelve-month campaign
  (impractically long at RHIC) may be performed in just three months (a long run, but feasible) with the EPD upgrade.

%%%%%%%%%%%%%%%%%%%%% Acknowledgements

\section*{Acknowledgements}

This work funded by U.S. National Science Foundation under Grants 1614835 
and 1614474, by the Ministry of Science and Technology (MoST) of China under grant No. 2016YFE0104800, and by 
the Office of Nuclear Physics within the U.S. DOE Office of Science.
\vspace*{2mm}\\
We gratefully acknowledge substantial support of
Elke Aschenauer,
Les Bland,
Tim Camarda, 
Mike Graham,
Leo Greiner,
Wlodek Guryn,
George Halal,
Will Jacobs,
Sammy Lisa,
Akio Ogawa,
Rahul Sharma,
Robert Soja,
Mikhail Stepanov,
Steve Valentino,
Flemming Videbaek,
Zhangbu Xu,
Lei Yang,
the LBNL carbon fiber lab,
the STAR Trigger group,
and the STAR Technical Support group.

\bibliographystyle{elsarticle-num}

\end{document}